\documentclass[twocolumn,amsmath,amssymb,aps]{revtex4-1}

\usepackage{graphicx}
\usepackage{dcolumn}
\usepackage{bm}
\usepackage{subfig}
\usepackage{natbib} 
\usepackage{multirow}
\usepackage{soul}
\usepackage{float}
\usepackage[normalem]{ulem}
\usepackage[usenames,dvipsnames]{color}

\newcommand{\bra}[1] { \langle #1 | }
\newcommand{\ket}[1] { | #1 \rangle }
\newcommand{\braket}[2] { \langle #1 | #2 \rangle }
\newcommand{\Bra}[1] { \Bigl\langle #1 \Bigr| }
\newcommand{\Ket}[1] { \Bigl| #1 \Bigr\rangle }

\newcommand{\nks}[0] { N_{\mathbf{k}} }

\newcommand{\nksinv}[0] { \frac{1}{N_{\mathbf{k}}} }
\newcommand{\nksinvv}[0] { \frac{1}{\sqrt{N_{\mathbf{k}}}} }
\newcommand{\nqs}[0] { N_{\mathbf{q}} }
\newcommand{\nqsinv}[0] { \frac{1}{N_{\mathbf{q}}} }

\newcommand{\nbnd}[0] { N_{\mathrm{b}} }

\def\QE{\textsc{Quantum ESPRESSO}\,}

\begin{document}

\captionsetup[subfigure]{labelformat=empty}

\title{Self-consistent Hubbard parameters from 
density-functional perturbation theory in the ultrasoft and 
projector-augmented wave formulations}

\author{Iurii Timrov,$^{1,*}$ Nicola Marzari,$^1$ and Matteo Cococcioni$^2$}
\affiliation{
$^1$Theory and Simulation of Materials (THEOS), and National Centre for Computational Design and Discovery of Novel Materials (MARVEL), \'Ecole Polytechnique F\'ed\'erale de Lausanne (EPFL), CH-1015 Lausanne, Switzerland\\
$^2$Department of Physics, University of Pavia, via Bassi 6, I-27100 Pavia, Italy\\
$^*$ e-mail: iurii.timrov@epfl.ch}

\date{\today}

\begin{abstract}
The self-consistent evaluation of Hubbard parameters using linear-response theory is crucial for quantitatively predictive calculations based on Hubbard-corrected density-functional theory. Here, we extend a recently-introduced approach based on density-functional perturbation theory (DFPT) for the calculation of the on-site Hubbard $U$ to also compute the inter-site Hubbard $V$. DFPT allows us to reduce significantly computational costs, improve numerical accuracy, and fully automate the calculation of the Hubbard parameters by recasting the linear response of a localized perturbation into an array of monochromatic perturbations that can be calculated in the primitive cell. In addition, here we generalize the entire formalism from norm-conserving to ultrasoft and projector-augmented wave formulations, and to metallic ground states. After benchmarking DFPT against the conventional real-space Hubbard linear response in a supercell, we demonstrate the effectiveness of the present extended Hubbard formulation in determining the equilibrium crystal structure of Li$_x$MnPO$_4$ ($x$=0,1) and the subtle energetics of Li intercalation.
\end{abstract}

\maketitle

\section{Introduction}
\label{sec:Introduction}

Exchange-correlation (xc) functionals based on the Hubbard extensions~\cite{anisimov:1991, anisimov:1997, dudarev:1998} to density-functional theory (DFT)~\cite{Hohenberg:1964,Kohn:1965} have proven to be quite effective in describing complex systems, both solid-state and molecular (see e.g. Refs.~\cite{Liechtenstein:1995, Kulik:2006, Kulik:2008, Zhou:2004, Huang:2008, Hsu:2009, Hampel:2017}). As it is well-known from the literature on DFT+$U$ and DFT with dynamical mean-field theory (DFT+DMFT)~\cite{Georges:1992, Georges:1996, anisimov:1997b, Liechtenstein:1998, Kotliar:2006}, a reliable and consistent method to evaluate the Hubbard parameters is key to achieving quantitative predictions. Although the empirical determination of these parameters is still common practice,
different methods to compute their value from first principles have been introduced, including constrained DFT (cDFT)~\cite{Dederichs:1984, Mcmahan:1988, Gunnarsson:1989, Hybertsen:1989, Gunnarsson:1990, Pickett:1998, Solovyev:2005, Nakamura:2006, Shishkin:2016}, Hartree-Fock based approaches~\cite{Mosey:2007, Mosey:2008, Andriotis:2010, Agapito:2015, TancogneDejean:2020, Lee:2020}, and the constrained random phase approximation (cRPA)~\cite{Springer:1998, 
Kotani:2000, Aryasetiawan:2004, Aryasetiawan:2006, Sasioglu:2011, Vaugier:2012, Amadon:2014, Seth:2017, Panda:2017, Nakamura:2021}. A linear-response formulation of cDFT (LR-cDFT) was introduced in Ref.~\cite{Cococcioni:2005} and generalized to the calculation of the inter-site Hubbard parameters $V$ in Ref.~\cite{Campo:2010} (see also Refs.~\cite{Kulik:2011, Himmetoglu:2014}). Recent work~\cite{Cococcioni:2019, Ricca:2020} involving some of the present authors, has highlighted the quantitative accuracy of DFT+$U$+$V$ calculations~\cite{Campo:2010} with both on-site $U$ and inter-site $V$ effective parameters (here, $V$ bridges the on-site Hubbard manifold with the surrounding ligand ions), when based on the self-consistent evaluation of the Hubbard parameters
using linear-response theory. 

LR-cDFT has recently been recast via density-functional perturbation theory (DFPT)~\cite{Timrov:2018}, allowing us to overcome several challenges of the supercell approach of Ref.~\cite{Cococcioni:2005}. 
In fact, by constructing the response of the system to a localized perturbation through a series of independent monochromatic perturbations to the primitive unit cell (rather than from finite-differences between calculations in supercells as in LR-cDFT), it improves significantly the computational efficiency, accuracy, user-friendliness, and automation ~\cite{Timrov:2018}, as also demonstrated by several recent applications~\cite{Cococcioni:2019,Ricca:2019,Ricca:2020,Floris:2020,Timrov:2020}. Key to this is indeed the capability to express perturbation theory in reciprocal space~\cite{Baroni:1987, Giannozzi:1991, Baroni:2001}. 
It is important to mention that the present formulation (be it in a LR-cDFT or DFPT implementation) aims to correct the over-delocalization and over-hybridization of the electrons in the localized Hubbard manifold; for this reason it is not appropriate to deal with closed-shell systems, where the electrons are fully contained in the localized manifold~\cite{Yu:2014}.

Encouraged by the significant improvement in flexibility and accuracy obtained with DFT+$U$+$V$ and spurred by the need to calculate Hubbard parameters efficiently and consistently, we present here a generalization of the DFPT implementation to compute also the inter-site Hubbard $V$ in a self-consistent fashion. In addition, we extend the formulation to metallic systems [i.e., with fractionally occupied Kohn-Sham (KS) states], to ultrasoft (US) pseudopotentials (PPs)~\cite{Vanderbilt:1990} and to the projector-augmented wave (PAW) method~\cite{Blochl:1994, Kresse:1999}. These extensions are particularly useful for systems in which transferability is paramount, magnetism arises, and localized valence states are present (e.g., semicore states promoted to valence or atomic states of the $d$ or $f$ kind in transition-metal and rare-earth compounds), which require the use of very high kinetic energy cut-offs in the plane-wave expansion. 

As a test case, we perform DFT+$U$ and DFT+$U$+$V$ calculations with self-consistent first-principles Hubbard parameters $U$ and $V$ (determined using DFPT) to study the structural properties and energetics (voltages) of a prototypical Li-ion cathode material Li$_x$MnPO$_4$ (x=0,1). Notably, the inclusion of the inter-site Hubbard $V$ improves the agreement of computed lattice parameters and voltages compared to the experimental values, as opposed to the case when only $U$ is taken into account.

The paper is organized as follows. Section~\ref{sec:DFT_Hub} summarizes the formulation of DFT+$U$+$V$~\cite{Campo:2010}; Sec.~\ref{ucalc} illustrates the linear-response approach to the calculation of Hubbard parameters $U$ and $V$, starting from a real-space formulation~\cite{Cococcioni:2005} (Sec.~\ref{ucalc_real}), reviewing its recent reciprocal-space implementation based on DFPT (Sec.~\ref{ucalc_rec}), and adapting the latter to US and PAW formulations (Sec.~\ref{sec:DFPT_HUB_monochromatic}); this is the main methodological advancement of the present paper. Section~\ref{sec:protocol} details a protocol designed to calculate the Hubbard parameters self-consistently, Sec.~\ref{sec:technical_details} contains the technical details of our calculations, and Sec.~\ref{sec:results}, after benchmarking the Hubbard parameters of MnPO$_4$ and LiMnPO$_4$ obtained from DFPT against those computed with LR-cDFT, introduces a self-consistent procedure for the evaluations of these parameters and discusses and contrasts the structural properties and average voltages (vs Li/Li$^+$) obtained using DFT+$U$ and DFT+$U$+$V$. Finally, Sec.~\ref{sec:Conclusions} presents our conclusive remarks. Some technical details regarding the implementation of DFT+$U$+$V$ with PAW and Bloch sums are presented in Appendices~\ref{app:PAW} and~\ref{app:Bloch_sums}, respectively. Hartree atomic units are used throughout the paper. For consistency, in this work we use notations as close as possible to those of Refs.~\cite{Timrov:2018, Dalcorso:2001}.

\section{Extended Hubbard functionals: DFT+$U$+$V$}
\label{sec:DFT_Hub}

In this section we briefly review the formulation of the extended DFT+$U$+$V$ approach with norm-conserving (NC) PPs  (introduced in Ref.~\cite{Campo:2010} and further discussed in Refs.~\cite{Himmetoglu:2014} and \cite{Cococcioni:2019}), and then we discuss its generalizations to the case of US PPs and the PAW method. This discussion will highlight the differences with the standard DFT+$U$ functional, with a particular attention to the Hubbard potential that enters the unperturbed KS Hamiltonian. This is, in fact, an important ingredient for DFPT calculations of the Hubbard parameters when a self-consistent evaluation is required.

As a generalization of DFT+$U$, DFT+$U$+$V$ is also based on an additive correction to the approximate DFT energy functional, modeled on the Hubbard Hamiltonian~\cite{Campo:2010}: 
\begin{equation}
E_{\mathrm{DFT}+U+V} = E_{\mathrm{DFT}} + E_{U+V} \,.
\label{eq:Edft_plus_u}
\end{equation}
Here, $E_{\mathrm{DFT}}$ represents the approximate DFT energy (constructed, e.g., within the local spin density approximation -- LSDA, or the spin-polarized generalized-gradient approximation -- GGA), while $E_{U+V}$ contains the additional Hubbard term. 
At variance with the DFT+$U$ approach, containing only on-site interactions, DFT+$U$+$V$ is based on the extended Hubbard model including also inter-site interactions between an atom and its surrounding ligands. As first described in Ref.~\cite{Campo:2010}, in the simplified rotationally-invariant formulation~\cite{dudarev:1998}, the extended Hubbard term reads:
\begin{eqnarray}
E_{U+V} & = & \frac{1}{2} \sum_I \sum_{\sigma m_1 m_2} 
U^I \left( \delta_{m_1 m_2} - n^{II \sigma}_{m_1 m_2} \right) n^{II \sigma}_{m_2 m_1} \nonumber \\
& & - \frac{1}{2} \sum_{I} \sum_{J (J \ne I)}^* \sum_{\sigma m_1 m_2} V^{I J} 
n^{I J \sigma}_{m_1 m_2} n^{J I \sigma}_{m_2 m_1} \,,
\label{eq:Edftu}
\end{eqnarray}
where $I$ and $J$ are atomic site indices, $m_1$ and $m_2$ are the magnetic quantum numbers associated with a specific angular momentum, $U^I$ and $V^{I J}$ are the on-site and inter-site Hubbard parameters, and the star in the sum denotes that for each atom $I$, the index $J$ covers all its neighbors up to a given distance (or up to a given shell). 

The atomic occupation matrices $n^{I J \sigma}_{m_1 m_2}$
are based on a generalized projection of the KS states on localized orbitals $\varphi^{I}_{m_1}(\mathbf{r})$ of neighbor atoms: 
\begin{equation}
n^{I J \sigma}_{m_1 m_2} = \sum_{\mathbf{k}}^{\nks} \sum_v 
\tilde{\theta}_{F, v\mathbf{k}\sigma} \, 
\bra{\psi^\circ_{v\mathbf{k}\sigma}} \hat{P}^{J I}_{m_2 m_1}
\ket{\psi^\circ_{v\mathbf{k}\sigma}} \,, 
\label{eq:occ_matrix_0}
\end{equation}
where $v$ and $\sigma$ represent, respectively, the band and spin labels of the KS states, $\mathbf{k}$ indicates points in the first Brillouin zone (BZ), $\nks$ is the number of $\mathbf{k}$ points in the first BZ, 
$\tilde{\theta}_{F, v\mathbf{k}\sigma}$ is the occupation of the KS states $\psi_{v\mathbf{k}\sigma}$~\cite{deGironcoli:1995, Timrov:Note:2020:ThetaOccupations} (to account for metals where these quantities can be fractional) and $\hat{P}^{J I}_{m_2 m_1}$ is the generalized projector on the localized orbitals of neighbor atoms:
\begin{equation}
\hat{P}^{J I}_{m_2 m_1} = 
\ket{\varphi^{J}_{m_2}} \bra{\varphi^{I}_{m_1}} \,.
\label{eq:Pm1m2}
\end{equation}
Here, $\varphi^I_{m_1}(\mathbf{r}) \equiv \varphi^{\gamma(I)}_{m_1}(\mathbf{r} - \mathbf{R}_I)$ are localized orbitals centered on the $I$th atom of type $\gamma(I)$ at the position $\mathbf{R}_I$. 
Given their importance for the calculation of the Hubbard parameters it is convenient to establish a short-hand notation for the on-site terms of the quantities defined in Eqs.~\eqref{eq:occ_matrix_0} and ~\eqref{eq:Pm1m2}:
\begin{equation}
n^{I\sigma}_{m_1m_2} \equiv n^{II\sigma}_{m_1m_2} \,,
\label{ndiag}
\end{equation}
and
\begin{equation}
\hat P^I_{m_1 m_2} \equiv \hat P^{II}_{m_1 m_2} \,.
\label{pdiag}
\end{equation}
The standard DFT+$U$ approach corresponds to the first line of Eq.~\eqref{eq:Edftu}. Based on the definitions above, it is quite straightforward to see from Eq.~\eqref{eq:Edftu} that the two terms in the corrective energy functional, proportional to the on-site ($U^{I}$) and inter-site ($V^{IJ}$) couplings, respectively, counteract each other. In fact, while the on-site term favors localization on atomic sites (typically suppressing hybridization), the inter-site one favors hybridized states with components on neighbor atoms. Computing the value of the $U^I$ and $V^{IJ}$ effective interaction parameters is thus crucial to determine the degree of atomic localization of $d$- and/or $f$-type electrons when the system is in its ground state. The Hubbard manifold $\{ \varphi^{I}_{m_1}(\mathbf{r}) \}$ can be constructed from the nonorthogonalized atomic orbitals (which are provided with PPs and which are orthonormal within each atom), or from the orthogonalized atomic orbitals (which are obtained by orthogonalizing the atomic orbitals from different sites)~\cite{Timrov:2020b}, or using (maximally localized) Wannier functions~\cite{Marzari:2012}.

For the purpose of this work, it is important to study the contribution to the KS potential stemming from the extended Hubbard correction. The action of this term on a KS pseudo-wavefunction can be easily
obtained by taking the functional derivative of $E_{\mathrm{DFT}+U+V}$ [see Eq.~\eqref{eq:Edft_plus_u}] with respect to the complex conjugate of the same pseudo-wavefunction~\cite{Campo:2010, Timrov:Note:2018:errorJPCM}. The term corresponding to this functional derivative of $E_{U+V}$ [see Eq.~\eqref{eq:Edftu}] is:
\begin{eqnarray}
\hat{V}^\circ_{\mathrm{Hub},\sigma} & = & 
\sum_I \sum_{m_1 m_2} U^{I} \left( \frac{\delta_{m_1 m_2}}{2} - 
n^{I \sigma}_{m_1 m_2} \right) \hat{P}^{I}_{m_1 m_2} \nonumber \\
& & - \sum_{I} \sum_{J (J \ne I)}^* \sum_{m_1 m_2} V^{I J} 
n^{I J \sigma}_{m_1 m_2} \hat{P}^{I J}_{m_1 m_2} \,.
\label{eq:Hub_pot_0}
\end{eqnarray}

In the following, we present the DFT+$U$+$V$ formalism for the case of US PPs and the PAW method. While many works are based on the projection of KS states on projector functions in the augmentation spheres (see Eq.~(6) of Ref.~\cite{Vanderbilt:1990}), this work 
uses projections on orthogonalized atomic orbitals, which are not restricted to the augmentation spheres. The main modifications that various terms undergo when US PPs or the PAW method is used are related to the fact that KS pseudo-wavefunctions are made ``soft" by smoothing their profiles in the core regions so that it becomes necessary to augment the sum of their squares with the missing core-localized ``hard" parts in order to calculate the proper charge density~\cite{Timrov:Note:2020:USPP_aug}: 
\begin{eqnarray}
& & \hspace{0.5cm}
\rho^\circ_\sigma(\mathbf{r}) = \sum_{\mathbf{k}}^{\nks} \sum_v \tilde{\theta}_{F, v\mathbf{k}\sigma} \, 
\vert\psi^\circ_{v\mathbf{k}\sigma}(\mathbf{r})\vert^2 \nonumber \\ 
& & \hspace{0.cm} 
+ \sum_{\mathbf{k}}^{\nks} \sum_v \tilde{\theta}_{F, v\mathbf{k}\sigma}
\sum_{I\mu\nu} Q^{\gamma(I)}_{\mu\nu}(\mathbf{r}-\mathbf{R}_I) \,
\braket{\psi^\circ_{v\mathbf{k}\sigma}}{\beta^{I}_\mu} 
\braket{\beta^{I}_\nu}{\psi^\circ_{v\mathbf{k}\sigma}} \,. \nonumber \\
& & 
\label{eq:density_0}
\end{eqnarray}
As evident from this equation, the augmentation of the charge density is realized by adding localized augmentation charge densities $Q^{\gamma(I)}_{\mu\nu}(\mathbf{r} - \mathbf{R}_I)$ (that pertain to the pseudopotential of the $I$th atom, of type $\gamma(I)$ at position $\mathbf{R}_I$) through a projection of KS states on appropriately constructed localized functions $\beta^{I}_\mu(\mathbf{r}) \equiv \beta^{\gamma(I)}_\mu(\mathbf{r} - \mathbf{R}_I)$ that vanish at and beyond an atom-specific core radius ~\cite{Vanderbilt:1990} and that are labeled by atomic-state indices $\mu$ and $\nu$. Being deprived of the hard part around the nuclei, the soft KS states must satisfy a generalized orthonormality condition:
\begin{equation}
\bra{\psi^\circ_{v\mathbf{k}\sigma}} \hat{S} \ket{\psi^\circ_{v'\mathbf{k}'\sigma'}} 
= \delta_{v v'} \delta_{\mathbf{k} \mathbf{k}'} \delta_{\sigma \sigma'} \,,
\label{eq:orthonormality_00}
\end{equation}
where $\hat S$ is an overlap operator
\begin{equation}
\hat{S} = 1 + \sum_{I\mu\nu} q^{\gamma(I)}_{\mu\nu} \, \ket{\beta^{I}_\mu} \bra{\beta^{I}_\nu} \,,
\label{eq:S_overlap}
\end{equation}
based on the same $\beta$ functions of Eq.~\eqref{eq:density_0} and on augmentation charges defined as:
\begin{equation}
q^{\gamma(I)}_{\mu\nu} = 
\int\limits_\mathrm{V} Q^{\gamma(I)}_{\mu\nu}(\mathbf{r}) \, d\mathbf{r} \,, 
\label{intq}
\end{equation}
with $V$ being the volume of the crystal. As a consequence, the soft KS states are calculated by solving the generalized KS equations~\cite{Vanderbilt:1990, Timrov:Note:2018:circ}:
\begin{equation}
\hat{H}^\circ_{\sigma} \ket{\psi^\circ_{v\mathbf{k}\sigma}} = 
\varepsilon^\circ_{v\mathbf{k}\sigma} \hat{S} \ket{\psi^\circ_{v\mathbf{k}\sigma}} \,,
\label{eq:KSeq_GS}
\end{equation} 
where $\varepsilon^\circ_{v\mathbf{k}\sigma}$ are the KS energies, and
\begin{equation}
\hat{H}^\circ_\sigma = \hat{H}^\circ_{\mathrm{DFT},\sigma} + \hat{V}^\circ_{\mathrm{Hub},\sigma} \,,
\label{eq:H_tot_GS}
\end{equation}
with $\hat{H}^\circ_{\mathrm{DFT},\sigma}$ being the DFT Hamiltonian (LSDA or spin-polarized GGA) 
and $\hat{V}^\circ_{\mathrm{Hub},\sigma}$ is the Hubbard potential given by Eq.~\eqref{eq:Hub_pot_0}.
The overlap operator of Eq.~\eqref{eq:S_overlap} has to be used also as the kernel of all the scalar products between pseudo KS wavefunctions, as already exemplified in the generalized orthonormality condition, Eq.~\eqref{eq:orthonormality_00}. 
The effective Hubbard corrective potential, Eq.~\eqref{eq:Hub_pot_0}, 
can thus be effectively obtained from the inclusion of the overlap operator in the expression of the projector on atomic states, Eq.~\eqref{eq:Pm1m2}:
\begin{equation}
\hat{P}^{J I}_{m_2 m_1} = 
\hat{S} 
\ket{\varphi^{J}_{m_2}} \bra{\varphi^{I}_{m_1}} 
\hat{S} 
\,.
\label{eq:Pm1m2us}
\end{equation}
This generalized expression of the projector is sufficient to obtain the
atomic occupations as indicated in Eq.~\eqref{eq:occ_matrix_0} using ``soft" (atomic and KS) wavefunctions and, consequently, the generalization of the Hubbard energy, Eq.~\eqref{eq:Edftu}, and potential, Eq.~\eqref{eq:Hub_pot_0}, to implementations based on US PPs. The same generalized expression of the Hubbard functional obtained for US PPs can actually be used also for PAW, once augmentation quantities are properly adapted~\cite{Kresse:1999,Audouze:2008} (see
Sec.~\ref{sec:DFPT_PAW}). This point is discussed in more detail in Appendix~\ref{app:PAW}.

\section{Calculation of Hubbard parameters from linear-response theory}
\label{ucalc}

\subsection{Definition of Hubbard parameters}
\label{ucalc_real}

The importance of the calculation of Hubbard parameters from first-principles using linear-response theory stems from the capability of the Hubbard corrections to remove the residual electronic self-interactions from approximate energy functionals, which manifests itself through an undesirable curvature of the total energy as a function of atomic occupations~\cite{Cococcioni:2005}. In fact, Hubbard effective interactions can be defined as the second derivatives of the total energy with respect to the total  occupation of a given atom, i.e. with respect to the trace of the occupation matrix defined in Eq.~\eqref{ndiag}~\cite{Cococcioni:2005}; with this definition, Hubbard corrections restore the desired piecewise linearity of the total energy. This can be achieved by perturbing the system with a shift in the potential acting on the Hubbard states of a given atom, $\Delta \hat{V}^J = \lambda^J \sum_m \hat{P}^{J}_{m m}$, and then computing the response of all the atomic occupations. Applying this to all the Hubbard atoms in the system allows us to construct the {\it bare} and {\it self-consistent} susceptibility matrices (obtained, in practical calculations, respectively at the beginning of the perturbed run and at its self-consistent convergence):
\begin{equation}
    (\chi_0)_{IJ} = \frac{dn^I_0}{d\lambda^J}, \qquad
    (\chi)_{IJ} = \frac{dn^I}{d\lambda^J} \,.
    \label{chi}
\end{equation}
From these, the effective Hubbard parameters can be readily obtained~\cite{Cococcioni:2005}:
\begin{equation}
U^I = \left( \chi_0^{-1} - \chi^{-1} \right)_{II} \,,
\label{eq:U_def}
\end{equation} 
\begin{equation}
V^{IJ} = \left( \chi_0^{-1} - \chi^{-1} \right)_{IJ} \,.
\label{eq:V_def}
\end{equation}
It is important to stress that the procedure outlined above is based on isolated perturbations; therefore, it requires the use of large supercells (whose size has to be increased until the convergence of $U^I$ and $V^{IJ}$ is reached) that makes the computational cost and numerical noise increase rapidly~\cite{Cococcioni:2005}. However, DFPT allows us to largely reduce these issues~\cite{Timrov:2018}.

\subsection{DFPT with norm-conserving PPs: Extension to $V$ and to metallic systems}
\label{ucalc_rec}

In order to overcome the difficulties alluded to in Sec.~\ref{ucalc_real} we have recently introduced an implementation of the LR approach outlined above using DFPT~\cite{Timrov:2018}. The scope of this section is to discuss the extension of this formalism to metallic systems and to the calculation of the inter-site interaction parameters V. For the sake of simplicity, this discussion is based on a NC PPs implementation; the generalization to US PPs and the PAW method is discussed in Sec.~\ref{sec:DFPT_HUB_monochromatic}. Within the framework of the DFPT implementation, the response of the KS wavefunctions to a small perturbation of the atomic potential [that induces a variation of the atomic occupations defined in Eq.~\eqref{chi}] is obtained as the solution of the perturbative problem resulting from a first-order variation of the KS equations:
\begin{eqnarray}
& &\left( \hat{H}^\circ_{
\sigma} - \varepsilon^\circ_{v\mathbf{k}\sigma} 
\, \right) \Ket{\frac{d\psi_{v\mathbf{k}\sigma}}{d\lambda^{J}}} \nonumber \\ 
& & \hspace{0.3cm} = - \biggl( \frac{d\hat{V}_{\mathrm{Hxc},\sigma}}{d\lambda^{J}} 
- \frac{d \varepsilon_{v\mathbf{k}\sigma}}{d\lambda^{J}}
+ \hat{V}_\mathrm{pert}^{J} \biggr) \ket{\psi^\circ_{v\mathbf{k}\sigma}} \,,
\label{eq:KS_lin_eq_q}
\end{eqnarray}
where $\hat{V}_\mathrm{pert}^{J} = \sum_m \hat{P}^{J}_{mm}$ is the perturbing potential, and $\frac{d\hat{V}_{\mathrm{Hxc},\sigma}}{d\lambda^{J}}$, $\frac{d\psi_{v\mathbf{k}\sigma}}{d\lambda^{J}}$, and $\frac{d \varepsilon_{v\mathbf{k}\sigma}}{d\lambda^{J}}$ are the response Hartree and xc (Hxc) potentials, the response KS wavefunctions, and the response KS energies, respectively~\cite{Timrov:2018}.  
As detailed in Refs.~\cite{Baroni:2001, Timrov:2018}, 
the problem has to be solved self-consistently because the response of the KS eigenvalues and of the Hxc potential appearing on the right-hand side of Eq.~\eqref{eq:KS_lin_eq_q} depend on the response of the KS wavefunctions, obtained from the solution of the perturbative problem in the equation above. Once convergence is achieved, the variation of the diagonal (with respect to atomic sites) atomic occupation matrices [that define the self-consistent susceptibility matrix in Eq.~\eqref{chi}] is obtained~\cite{Timrov:Note:2020:nij}: 
\begin{eqnarray}
\frac{dn^{I \sigma}_{m_1 m_2}}{d\lambda^{J}} & = &
\sum_{\mathbf{k}}^{\nks} \sum_v \tilde{\theta}_{F, v\mathbf{k}\sigma}
\biggl[ \Bra{\psi^\circ_{v\mathbf{k}\sigma}} \hat{P}^{I}_{m_2 m_1} 
\Ket{\frac{d{\psi}_{v\mathbf{k}\sigma}}{d\lambda^{J}}} \biggr. \nonumber \\
& & \hspace{1.3cm} \biggl. + \, \Bra{\frac{d{\psi}_{v\mathbf{k}\sigma}}{d\lambda^{J}}} 
\hat{P}^{I}_{m_2 m_1} \Ket{\psi^\circ_{v\mathbf{k}\sigma}} \biggr] \nonumber \\ 
& & \hspace{-0.3cm} + \sum_{\mathbf{k}}^{\nks} \sum_v
\frac{d \tilde{\theta}_{F, v\mathbf{k}\sigma}}{d\lambda^J} \, \bra{\psi^\circ_{v\mathbf{k}\sigma}} \hat{P}^{I}_{m_2 m_1}
\ket{\psi^\circ_{v\mathbf{k}\sigma}} \,.
\label{eq:occ_matrix_response_new}
\end{eqnarray}
The last term on the right-hand side of Eq.~\eqref{eq:occ_matrix_response_new} comes from the derivative of the occupations of KS states appearing in Eq.~\eqref{eq:occ_matrix_0}~\cite{Dalcorso:2001}:
\begin{equation}
\frac{d\tilde{\theta}_{F, v\mathbf{k}\sigma}}{d\lambda^{J}} =
\frac{1}{\eta} \, \tilde{\delta} \Bigl( \frac{\varepsilon_F - \varepsilon^\circ_{v\mathbf{k}\sigma}}{\eta} \Bigr) \left[ \frac{d\varepsilon_F}{d\lambda^{J}} - \frac{d\varepsilon_{v\mathbf{k}\sigma}}{d\lambda^{J}} \right] \,,
\label{eq:dtheta}
\end{equation}
where $\tilde{\delta}(\varepsilon) \equiv d\tilde{\theta}(\varepsilon)/d\varepsilon$ is a smooth approximation to the Dirac's $\delta$ function, $\eta$ is a broadening parameter, and $\varepsilon_F$ and $\frac{d\varepsilon_F}{d\lambda^{J}}$ 
are the Fermi energy and its shift (see Sec.~II.C.4 of Ref.~\onlinecite{Baroni:2001}).
Obviously, the term $\frac{d\tilde{\theta}_{F, v\mathbf{k}\sigma}}{d\lambda^{J}}$ is absent for semiconductors and insulators as their KS state occupations are either 0 or 1. 

It is important to remark that, when LR calculations are based on the
DFT+$U$+$V$ ground state (for self-consistent evaluations of the Hubbard parameters~\cite{Cococcioni:2019,Ricca:2020,Timrov:2020}), the Hubbard correction enters the unperturbed Hamiltonian
$\hat{H}^\circ_{\sigma}$ in Eq.~\eqref{eq:KS_lin_eq_q},
with the Hubbard potential given in Eq.~\eqref{eq:Hub_pot_0}. 
However, as was pointed out in Ref.~\cite{Timrov:2018}, 
the response of the Hubbard potential is not present in Eq.~\eqref{eq:KS_lin_eq_q}
so that the Hubbard parameters are obtained, consistently with their definition, as second derivatives of the DFT part only of the total energy.

The major advantage offered by the DFPT reformulation of LR-cDFT consists in the possibility to obtain the variation of atomic occupations as a sum of wavevector-specific contributions that can be computed independently from one another (thus leading to better scaling of the computational cost~\cite{Timrov:2018}) using the primitive cell of the system. In fact, the Fourier spectrum of a 
perturbation that has the periodicity of a supercell (as needed to eliminate the interactions with  periodic replicas) only contains fundamental vectors of its reciprocal lattice that map into a corresponding $\mathbf{q}$ points grid
within the Brillouin zone corresponding to the primitive cell~\cite{Timrov:2018}. The total response of atomic occupations can thus be written as (see Eq.~(42) 
in Ref.~\cite{Timrov:2018}):
\begin{equation}
\frac{dn^{s l \sigma}_{m_1 m_2}}{d\lambda^{s' l'}} =
\nqsinv \sum_\mathbf{q}^{\nqs} e^{i\mathbf{q}\cdot(\mathbf{R}_{l}-\mathbf{R}_{l'})} \, 
\Delta^{s'}_\mathbf{q} \bar{n}^{s \, \sigma}_{m_1 m_2} \,,
\label{eq:occ_matrix_response_lp_1}
\end{equation}
where the atomic site indices $I$ and $J$ have been decomposed as $I = (l,s)$ and $J = (l',s')$ indicating, respectively, the cell the atom belongs to ($l$ and $l'$) and its position within the cell ($s$ and $s'$). Here, $N_\mathbf{q}$ is the number of $\mathbf{q}$ points in the first BZ (note that the dimension of the $\mathbf{q}$ points grid reflects directly the dimension of the supercell of which it is the reciprocal-space image).
$\Delta^{s'}_\mathbf{q} \bar{n}^{s \, \sigma}_{m_1 m_2}$ represents the lattice-periodic response (hence the over-bar) of the occupation matrix to a monochromatic perturbation with a wavevector $\mathbf{q}$, and it can be linked to the lattice-periodic variations of the KS wavefunctions as follows \cite{Timrov:2018}:
\begin{eqnarray}
\Delta^{s'}_\mathbf{q} \bar{n}^{s \, \sigma}_{m_1 m_2} & = &
\nksinv \sum_{\mathbf{k}}^{\nks} \sum_{v}^{\nbnd}
\biggl[ \Bra{\bar{u}^\circ_{v\mathbf{k}\sigma}} 
\hat{\bar{P}}^{s}_{m_2,m_1,\mathbf{k},\mathbf{k}+\mathbf{q}}
\Ket{\Delta^{s'}_\mathbf{q} \bar{u}_{v\mathbf{k}\sigma}} \biggr. \nonumber \\ [4pt]
& & \hspace{0.6cm} + \, \biggl. \Bra{\bar{u}^\circ_{v\mathbf{k}\sigma}} 
\hat{\bar{P}}^{s}_{m_1,m_2,\mathbf{k},\mathbf{k}+\mathbf{q}}
\Ket{\Delta^{s'}_\mathbf{q} \bar{u}_{v\mathbf{k}\sigma}} \biggr] \nonumber \\
& & \hspace{-1.5cm} + \delta_{\mathbf{q}, \mathbf{0}} \, \nksinv \sum_{\mathbf{k}}^{\nks} \sum_v^{\nbnd} 
\frac{1}{\eta} \, \tilde{\delta}\left(\frac{\varepsilon_F - 
\varepsilon^\circ_{v\mathbf{k}\sigma}}{\eta}\right) \Delta^{s'}_\mathbf{0} \varepsilon_{F} \, 
\nonumber \\ [4pt]
& & \hspace{0.3cm} \times \, \bra{\bar{u}^\circ_{v\mathbf{k}\sigma}} 
\hat{\bar{P}}^{s}_{m_2,m_1,\mathbf{k},\mathbf{k}}
\ket{\bar{u}^\circ_{v\mathbf{k}\sigma}} \,.
\label{eq:occ_matrix_response_lp_2}
\end{eqnarray}
Here, $\bar{u}^\circ_{v\mathbf{k}\sigma}$ and $\Delta^{s'}_\mathbf{q} \bar{u}_{v\mathbf{k}\sigma}$
are the lattice-periodic parts of the unperturbed and linear-response monochromatic $\mathbf{q}$ component of the KS wavefunctions, respectively (see the appendices~1 and 3 in Ref.~\cite{Timrov:2018}). In the derivation of Eq.~\eqref{eq:occ_matrix_response_lp_2}, we used the expression $\frac{d\varepsilon_F}{d\lambda^{s' l'}} = \frac{1}{N_\mathbf{q}} \Delta^{s'}_\mathbf{0} \varepsilon_{F}$, where $\Delta^{s'}_\mathbf{0} \varepsilon_{F}$ is the shift of the Fermi energy that occurs when applying a macroscopic perturbation (i.e. $\mathbf{q} = \boldsymbol{0}$) to the $s'$th atom in the reference primitive cell~\cite{Timrov:Note:2018:dEf}. The lattice-periodic part of the projector on the Hubbard manifold, which appears in Eq.~\eqref{eq:occ_matrix_response_lp_2}, reads~\cite{Timrov:2018}:
\begin{equation}
\hat{\bar{P}}^{
s}_{m_2,m_1,\mathbf{k},\mathbf{k}} = 
\ket{\bar{\varphi}^{s}_{m_2,\mathbf{k}}}
\bra{\bar{\varphi}^s_{m_1,\mathbf{k}}} \,.
\label{eq:P_proj_lp0}
\end{equation}
The functions $\bar{\varphi}^s_{m_1,\mathbf{k}}$ are defined in the same way as in Appendix~1 of Ref.~\cite{Timrov:2018}.
The first two terms on the right-hand side of Eq.~\eqref{eq:occ_matrix_response_lp_2} were made to look similar (except for the inversion in the order of indices $m_1$ and $m_2$) by the use of time-reversal symmetry. As was mentioned above, due to the linear character of the perturbative problem [Eq.~\eqref{eq:KS_lin_eq_q}], the lattice-periodic components of the response KS wavefunctions at different $\mathbf{q}$ can be computed independently from one another as solutions of $\mathbf{q}$-specific Sternheimer equations~\cite{Timrov:2018}:
\begin{eqnarray}
& & \left( \hat{\bar{H}}^\circ
_{\mathbf{k+q},\sigma} + 
\alpha \hat{\bar{\mathcal{O}}}_{\mathbf{k+q},\sigma} 
- \varepsilon^\circ_{v\mathbf{k}\sigma} \right) \, 
\ket{\Delta^{s'}_\mathbf{q} \bar{u}_{v\mathbf{k}\sigma}} \nonumber \\
& & \hspace{0.4cm} = - \hat{\bar{\mathcal{P}}}^\dagger_{v,\mathbf{k},\mathbf{k+q}, \sigma}
\left( \, \Delta^{s'}_\mathbf{q} 
\hat{\bar{V}}_{\mathrm{Hxc},\sigma} +
\hat{\bar{V}}^{s'}_{\mathrm{pert}, \mathbf{k+q}, \mathbf{k}} \right) \, 
\ket{\bar{u}^\circ_{v\mathbf{k}\sigma}} \,, \nonumber \\
& & 
\label{eq:LRKSeq_lp}
\end{eqnarray}
where the perturbing potential is constructed from a generalization of the projector in Eq.~\eqref{eq:P_proj_lp0}:
\begin{equation}
\hat{\bar{V}}^{s'}_{\mathrm{pert}, \mathbf{k}+\mathbf{q},\mathbf{k}} = 
\sum_m \hat{\bar{P}}^{s'}_{m,m,\mathbf{k}+\mathbf{q},\mathbf{k}}\,.
\label{vpert}
\end{equation}
The quantities $\hat{\bar{H}}^\circ_{\mathbf{k+q},\sigma}$ and $\Delta^{s'}_\mathbf{q} \hat{\bar{V}}_{\mathrm{Hxc},\sigma}$ are, respectively, the lattice-periodic parts of the unperturbed Hamiltonian $\hat{H}^\circ_\sigma$ (which contains the Hubbard corrective potential with on-site $U$ and inter-site $V$) and the response Hxc potential for a specific $\mathbf{q}$. The latter quantity depends on the response spin charge density at the same $\mathbf{q}$.
The operators $\hat{\bar{\mathcal{O}}}_{\mathbf{k+q},\sigma}$ and $\hat{\bar{\mathcal{P}}}^\dagger_{v,\mathbf{k}, \mathbf{k+q}, \sigma}$
are the lattice-periodic parts of generalized projectors on the valence and conduction manifolds, respectively~\cite{Baroni:2001,Dalcorso:2001}:
\begin{equation}
\hat{\bar{\mathcal{O}}}_{\mathbf{k+q},\sigma} =  \sum_{v'}^{\nbnd} 
\ket{\bar{u}^\circ_{v'\mathbf{k+q}\sigma}} \bra{\bar{u}^\circ_{v'\mathbf{k+q}\sigma}} \,,
\label{eq:proj_v}
\end{equation}
and
\begin{equation}
\hat{\bar{\mathcal{P}}}^\dagger_{v,\mathbf{k},\mathbf{k+q},\sigma} = 
\tilde{\theta}_{F, v\mathbf{k}\sigma} - 
\sum_{v'}^{\nbnd} \gamma_{v\mathbf{k}\sigma, \, v'\mathbf{k+q}\sigma} \, 
\ket{\bar{u}^\circ_{v'\mathbf{k+q}\sigma}} \bra{\bar{u}^\circ_{v'\mathbf{k+q}\sigma}} \,,
\label{eq:proj_c}
\end{equation}
with
\begin{eqnarray}
\gamma_{v\mathbf{k}\sigma, \, v'\mathbf{k+q}\sigma} & = & 
\tilde{\theta}_{F, v\mathbf{k}\sigma} \, \theta_{v\mathbf{k}\sigma, \, v'\mathbf{k+q}\sigma} + 
\tilde{\theta}_{F, v'\mathbf{k+q}\sigma} \, \theta_{v'\mathbf{k+q}\sigma, \, v\mathbf{k}\sigma} \nonumber \\
& & +  \, \alpha \, \frac{\tilde{\theta}_{F, v\mathbf{k}\sigma} - 
\tilde{\theta}_{F, v'\mathbf{k+q}\sigma}}{\varepsilon^\circ_{v\mathbf{k}\sigma} - 
\varepsilon^\circ_{v'\mathbf{k+q}\sigma}} \, 
\theta_{v'\mathbf{k+q}\sigma, \, v\mathbf{k}\sigma} \,.
\label{gamma}
\end{eqnarray}
In Eqs.~\eqref{eq:LRKSeq_lp} and \eqref{gamma}, $\alpha = \varepsilon_F + 3 \eta - 
\mathrm{min}[\varepsilon^\circ_{v\mathbf{k}\sigma}]$, where 
$\mathrm{min}[\varepsilon^\circ_{v\mathbf{k}\sigma}]$ is the lowest KS energy at the considered $\mathbf{k}$ point, and $\theta_{v\mathbf{k}\sigma, \, v'\mathbf{k}'\sigma} \equiv \theta[(\varepsilon^\circ_{v\mathbf{k}\sigma} - \varepsilon^\circ_{v'\mathbf{k}'\sigma})/\eta]$ is a (smooth) step-like function chosen to be equal to the rescaled complementary error function, $\theta(\varepsilon) \equiv \mathrm{erfc}(-\varepsilon)/2$~\cite{Timrov:Note:2017:theta}. 
The summation over the electronic band index $v'$ in Eqs.~\eqref{eq:proj_v} and \eqref{eq:proj_c} (and over $v$ in Eq.~\eqref{eq:occ_matrix_response_lp_2}) is limited to the (either full or partially occupied) $\nbnd$ states that are 
below $\varepsilon_F + 3\eta$~\cite{deGironcoli:1995}, thus conveniently avoiding summations over the entire empty-states manifold.
The operator $\hat{\bar{\mathcal{O}}}_{\mathbf{k+q},\sigma}$ is inserted on the left-hand side of Eq.~\eqref{eq:LRKSeq_lp} in order to avoid singularity issues during the iterative solution; at the same time the operator $\hat{\bar{\mathcal{P}}}^\dagger_{v,\mathbf{k},\mathbf{k+q},\sigma}$ avoids very expensive sums over numerous empty states~\cite{Baroni:2001, Timrov:2018, Timrov:Note:2020:Pproj}. Note that due to the presence of the projector $\hat{\bar{\mathcal{P}}}^\dagger_{v,\mathbf{k},\mathbf{k+q},\sigma}$ in Eq.~\eqref{eq:LRKSeq_lp} the derivative of the KS eigenvalues disappears from the right-hand side of Eq.~\eqref{eq:LRKSeq_lp} in comparison to Eq.~\eqref{eq:KS_lin_eq_q} (see also Ref.~\cite{Timrov:2018}). All the operators in Eq.~\eqref{eq:LRKSeq_lp} appear with a specific $\mathbf{q}$ component as a result of
recasting Eq.~\eqref{eq:KS_lin_eq_q} in reciprocal space through the Bloch sums of all the quantities appearing in there (this is discussed in detail in Ref.~\cite{Timrov:2018}). The potential terms appearing on the right-hand side of Eq.~\eqref{eq:LRKSeq_lp} represent the lattice-periodic components of the corresponding potential variations at the indicated wavevector~$\mathbf{q}$. 
Once these equations are solved (self-consistently) for all the wavevectors, Eqs.~\eqref{eq:occ_matrix_response_lp_1} and \eqref{eq:occ_matrix_response_lp_2} are used to compute the susceptibility matrices using Eq.~\eqref{chi}, from which the Hubbard interaction parameters are readily obtained as indicated in Eqs.~\eqref{eq:U_def} and \eqref{eq:V_def}.

\subsection{Extension of DFPT to ultrasoft PPs and PAW}
\label{sec:DFPT_HUB_monochromatic}

After reviewing the DFPT formalism for the calculation of the Hubbard parameters, we specialize it here to US PPs and the PAW method. We stress that only Hubbard related terms will be discussed. The reader who is interested in a general discussion of DFPT with US PPs or PAW is encouraged to review the existing literature on the topic~\cite{Dalcorso:2001, Dalcorso:2010}.

\subsubsection{Ultrasoft pseudopotentials}

The Sternheimer equations that are solved in DFPT must be generalized to the US PPs case. After extensive but straightforward mathematical manipulations it can be shown that the final form of Eq.~\eqref{eq:LRKSeq_lp} in the US PPs case becomes:
\begin{eqnarray}
& & \left( \hat{\bar{H}}^\circ_{\mathbf{k+q},\sigma} + 
\alpha \hat{\bar{\mathcal{O}}}_{\mathbf{k+q},\sigma} 
- \varepsilon^\circ_{v\mathbf{k}\sigma} \, 
\hat{\bar{S}}_{\mathbf{k+q}} 
\right) \, 
\ket{\Delta^{s'}_\mathbf{q} \bar{u}_{v\mathbf{k}\sigma}} \nonumber \\
& & \hspace{0.4cm} = - \hat{\bar{\mathcal{P}}}^\dagger_{v,\mathbf{k}, \mathbf{k+q}, \sigma} 
\left( \, \Delta^{s'}_\mathbf{q} \hat{\bar{V}}_{\mathrm{eff},\mathbf{k+q},\mathbf{k},\sigma} + 
\hat{\bar{V}}^{s'}_{\mathrm{pert}, \mathbf{k+q}, \mathbf{k}} \right) \, 
\ket{\bar{u}^\circ_{v\mathbf{k}\sigma}} \,, \nonumber \\
& & 
\label{eq:LRKSeq_lp_us}
\end{eqnarray}
where $\hat{\bar{\mathcal{O}}}_{\mathbf{k+q},\sigma}$ and $\hat{\bar{\mathcal{P}}}^\dagger_{v,\mathbf{k},\mathbf{k+q},\sigma}$ are generalized as~\cite{Timrov:Note:2020:extraS}
\begin{equation}
\hat{\bar{\mathcal{O}}}_{\mathbf{k+q},\sigma} =  \sum_{v'}^{\nbnd} \hat{\bar{S}}_{\mathbf{k+q}}
\ket{\bar{u}^\circ_{v'\mathbf{k+q}\sigma}} \bra{\bar{u}^\circ_{v'\mathbf{k+q}\sigma}} \hat{\bar{S}}_{\mathbf{k+q}} \,,
\label{eq:proj_v0}
\end{equation}
and
\begin{eqnarray}
\hat{\bar{\mathcal{P}}}^\dagger_{v,\mathbf{k},\mathbf{k+q},\sigma} & = &  
\tilde{\theta}_{F, v\mathbf{k}\sigma} - 
\sum_{v'}^{\nbnd} \gamma_{v\mathbf{k}\sigma, \, v' \mathbf{k+q} \sigma} \nonumber \\
& & \hspace{0.5cm}\times \, \hat{\bar{S}}_{\mathbf{k+q}}
\ket{\bar{u}^\circ_{v'\mathbf{k+q}\sigma}} \bra{\bar{u}^\circ_{v'\mathbf{k+q}\sigma}} \,.
\label{eq:proj_c0}
\end{eqnarray}
In Eqs.~\eqref{eq:LRKSeq_lp_us}--\eqref{eq:proj_v0}, $\hat{\bar{S}}_\mathbf{k+q}$ is the lattice-periodic part of the overlap operator $\hat{S}$, which reads
\begin{equation}
\hat{\bar{S}}_\mathbf{k+q} = 1 + 
\sum_{s\mu\nu} q^{\gamma(s)}_{\mu\nu} \, \ket{\bar{\beta}^s_{\mu, \mathbf{k+q}}} 
\bra{\bar{\beta}^s_{\nu, \mathbf{k+q}}} \,,
\label{eq:S_k_plus_q}
\end{equation}
and the orthonormality condition~\eqref{eq:orthonormality_00} now reads $\bra{\bar{u}^\circ_{v\mathbf{k}\sigma}} \hat{\bar{S}}_\mathbf{k} 
\ket{\bar{u}^\circ_{v'\mathbf{k}\sigma'}} = \delta_{v v'} \delta_{\sigma \sigma'}$,
where the inner product is computed via integration over the volume of the primitive unit cell.
In Eq.~\eqref{eq:LRKSeq_lp_us}, the effective response potential $\Delta^{s'}_\mathbf{q} \hat{\bar{V}}_{\mathrm{eff},\mathbf{k+q},\mathbf{k},\sigma}$ is defined as a sum of the standard Hxc response potential $\Delta^{s'}_\mathbf{q} \hat{\bar{V}}_{\mathrm{Hxc},\sigma}$ [as in Eq.~\eqref{eq:LRKSeq_lp}, but which depends on the US PPs augmented response spin charge density~\cite{Dalcorso:2001}] plus an extra US PPs specific term, namely:
\begin{eqnarray}
    \Delta^{s'}_\mathbf{q} \hat{\bar{V}}_{\mathrm{eff},\mathbf{k+q},\mathbf{k},\sigma} & = & 
    \Delta^{s'}_\mathbf{q}  \hat{\bar{V}}_{\mathrm{Hxc},\sigma} + 
    \sum_{s\mu\nu} \langle \bar{Q}^s_{\mu\nu,\mathbf{q}} | \Delta^{s'}_\mathbf{q} \bar{V}_{\mathrm{Hxc},\sigma} \rangle \nonumber \\
    & & \hspace{2cm} \times \, |\bar{\beta}^s_{\mu,\mathbf{k+q}}\rangle \langle \bar{\beta}^s_{\nu,\mathbf{k}}| \,.
    \label{eq:dVeffUS}
\end{eqnarray}
The quantities appearing in Eqs.~\eqref{eq:S_k_plus_q} and \eqref{eq:dVeffUS} are defined in Appendix~\ref{app:Bloch_sums}.

Various terms of the Hubbard correction undergo a similar generalization when US PPs are used.
The Hubbard potential that is included in the unperturbed Hamiltonian $\hat{\bar{H}}^\circ_{\mathbf{k+q},\sigma}$ of Eq.~\eqref{eq:LRKSeq_lp_us},  has the following 
expression: 
\begin{eqnarray}
& & \hat{\bar{V}}^\circ_{\mathrm{Hub},\mathbf{k+q},\sigma} = \nonumber \\
& & \hspace{0.5cm} \sum_s \sum_{m_1 m_2} U^{s} \left( \frac{\delta_{m_1 m_2}}{2} - n^{s \, \sigma}_{m_1 m_2} \right) 
\hat{\bar{P}}^{s}_{m_1, m_2, \mathbf{k+q}, \mathbf{k+q}} \nonumber \\ [6pt]
& & \hspace{0.3cm} - \sum_s \sum_{s' (s' \ne s)}^* \sum_{\mathbf{R}_{l'} - \mathbf{R}_{l}}^* \sum_{m_1 m_2} V^{s \, l \, s' l'} n^{s \, l \, s' l' \sigma}_{m_1 m_2} \nonumber \\ 
& & \hspace{3.5cm} \times \, \hat{\bar{P}}^{s \, l \, s' \, l'}_{m_1, m_2, \mathbf{k+q}, \mathbf{k+q}} \,,
\label{eq:Hub_pot_0_lp}
\end{eqnarray}
where
\begin{equation}
n_{m_1 m_2}^{s \, l \, s' l' \sigma} =
\nksinv \sum_{\mathbf{k}}^{\nks} \sum_v \tilde{\theta}_{F, v\mathbf{k}\sigma} \,
\bra{\bar{u}^\circ_{v\mathbf{k}\sigma}} \hat{\bar{P}}^{s' l' s\, l}_{m_2,m_1,\mathbf{k},\mathbf{k}}
\ket{\bar{u}^\circ_{v\mathbf{k}\sigma}} \,.
\label{eq:occ_matrix0_lp}
\end{equation}
In analogy with Eq.~\eqref{eq:proj_v0}, the periodic, $\mathbf{k}$-specific projector is generalized as follows:
\begin{equation}
\hat{\bar{P}}^{s' l' s \, l}_{m_2,m_1,\mathbf{k},\mathbf{k}} = 
e^{i \mathbf{k} \cdot \left( \mathbf{R}_{l} - \mathbf{R}_{l'} \right)} \,
\hat{\bar{S}}_{\mathbf{k}} \ket{\bar{\varphi}^{s'}_{m_2,\mathbf{k}}}
\bra{\bar{\varphi}^s_{m_1,\mathbf{k}}} \hat{\bar{S}}_{\mathbf{k}} \,.
\label{eq:P_proj_lp}
\end{equation}
where the phase factor accounts for the possibility that the two atomic wave functions to belong to neighbor cells. In the first term of Eq.~\eqref{eq:Hub_pot_0_lp}, $n^{s \, \sigma}_{m_1 m_2}$ corresponds to 
$n^{s \, l \, s \, l \, \sigma}_{m_1 m_2}$, and $\bar{P}^{s}_{m_1, m_2, \mathbf{k}, \mathbf{k}}$ corresponds to $\bar{P}^{s \, l \, s \, l}_{m_1, m_2, \mathbf{k}, \mathbf{k}}$ (note, $n^{s \, \sigma}_{m_1 m_2}$ and $\bar{P}^{s}_{m_1, m_2, \mathbf{k}, \mathbf{k}}$ do not depend on the index $l$). In Eq.~\eqref{eq:Hub_pot_0_lp} $s$ and $s'$ run over the number of atoms in the cell they belong to ($l$ and $l'$, respectively), and the sum over $\mathbf{R}_{l'} - \mathbf{R}_{l}$ runs over the number of cells (including the original cell, i.e. $l'=l$) that are constructed to generate pairs of sites interacting with 
the inter-site Hubbard $V^{s \, l \, s' l'}$ parameter.

An analogous generalization also has to be applied to the (site-diagonal) perturbing potential, Eq.~\eqref{vpert}, appearing on the right-hand side of Eq.~\eqref{eq:LRKSeq_lp_us}.
Its expression now corresponds to the trace of a projector that generalizes 
Eq.~\eqref{eq:P_proj_lp} (with $s'=s$) to the case with $\mathbf{q} \neq {\boldsymbol 0}$:
\begin{equation}
\hat{\bar{P}}^s_{m_2,m_1,\mathbf{k+q},\mathbf{k}} =
\hat{\bar{S}}_{\mathbf{k+q}} \ket{\bar{\varphi}^s_{m_2,\mathbf{k+q}}}
\bra{\bar{\varphi}^s_{m_1,\mathbf{k}}} \hat{\bar{S}}_{\mathbf{k}} \,.
\label{eq:P_proj_lp_q}
\end{equation}

Finally, the same generalized expression of the lattice-periodic projector on atomic states has to be used in the calculation of monochromatic, $\mathbf{q}$-dependent components of the linear-response occupation matrices, Eq.~\eqref{eq:occ_matrix_response_lp_2}, which are then employed to calculate the values of the Hubbard $U$ and $V$ parameters using Eqs.~\eqref{chi} - \eqref{eq:V_def}.

\subsubsection{PAW}
\label{sec:DFPT_PAW}

The calculation of the Hubbard parameters with the PAW method is very similar to the one reviewed above for US PPs. This is so because there are many similarities between US and PAW formalisms, as discussed in detail in Refs.~\cite{Kresse:1999, Audouze:2008}. However, there are a few points that must be clarified. All the equations presented above for US PPs are valid also in the case of PAW, after taking into account the following changes: $i)$~The augmentation functions $Q^{\gamma(I)}_{\mu\nu}$ and projector functions $\beta^I_\mu$ must be replaced by the ones from the PAW method, $ii)$~in Eq.~\eqref{eq:LRKSeq_lp_us} the terms $\hat{\bar{H}}^\circ_{\mathbf{k+q},\sigma}$ and $\Delta^{s'}_\mathbf{q} \hat{\bar{V}}_{\mathrm{eff},\mathbf{k+q},\mathbf{k},\sigma}$ have extra contributions (from standard DFPT with PAW) that are defined inside of the augmentation spheres centered about atoms~\cite{Dalcorso:2010}.

It is important to remark that our implementation is different from most implementations for DFT+$U$ with PAW (see, e.g., Refs.~\cite{Bengone:2000, Rohrbach:2003}), because our formalism is based on projecting KS wavefunctions on (orthogonalized) atomic orbitals $\varphi^I$ rather than on projector functions $\beta^I$ (the latter being localized inside the augmentation spheres). One advantage of our formalism is that it allows DFT+$U$+$V$ to be extended easily to PAW, [Eqs.~\eqref{eq:Hub_pot_0} and \eqref{eq:Pm1m2us}].

\section{Self-consistent calculation of Hubbard parameters}
\label{sec:protocol}

The increased computational efficiency, the higher level of automation, and the user-friendliness promoted by the use of DFPT in the calculation of the Hubbard parameters~\cite{Timrov:2018} can also be exploited to make this calculation fully ``self-consistent" and to obtain Hubbard parameters that are fully consistent with both the electronic structure of the system and, in addition, with the crystal structure. The idea is to recompute the effective $U$ and $V$ parameters from a DFT+$U$+$V$ ground state, until the values obtained from the DFPT calculation ($U_\mathrm{out}$ and $V_\mathrm{out}$) coincide (within a fixed precision $\Delta$) with those used in determining the ground state that DFPT is based on ($U_\mathrm{in}$ and $V_\mathrm{in}$):
\begin{equation}
    |U_\mathrm{out} - U_\mathrm{in}| < \Delta \,,
    \label{eq:U_conv}
\end{equation}
\begin{equation}
    |V_\mathrm{out} - V_\mathrm{in}| < \Delta \,.
    \label{eq:V_conv}
\end{equation}
The convergence (typically in a few cycles) of the iterative procedure in most cases is guaranteed by the observed smooth and monotonic dependence of output interaction parameters ($U_\mathrm{out}$, $V_\mathrm{out}$) on the input ones ($U_\mathrm{in}$ and $V_\mathrm{in}$). This iterative procedure yields the final ``self-consistent" values of the Hubbard interactions, labeled $U_\mathrm{scf}$ and $V_\mathrm{scf}$.

The idea to compute the Hubbard parameters self-consistently has been explored several times in the literature. The first attempt, to the best of our knowledge, was made in Ref.~\cite{Kulik:2006} (also involving some of us) where an extrapolation of the linear behavior of $U_\mathrm{out}$ (computed from LR-cDFT~\cite{Cococcioni:2005}) as a function of $U_\mathrm{in}$ was used to determine the self-consistent value of the Hubbard $U$ through the elimination of the second derivative of the Hubbard correction, consistently with the idea that $U_\mathrm{scf}$ should correspond to the curvature of the DFT part only of the total energy. In later works~\cite{Hsu:2009, Karlsson:2010, Campo:2010, Himmetoglu:2014, Agapito:2015, Shishkin:2016, Cococcioni:2019, TancogneDejean:2020, Lee:2020} the self-consistent evaluation of the Hubbard parameters was based on an iterative recalculation of their values, each step using the parameters determined at the previous one (or a linear combination of those computed at previous steps). The use of LR-cDFT (or DFPT) to re-calculate iteratively the Hubbard parameters until self-consistency is straightforward. However, in order to prevent the Hubbard correction from responding to the perturbation, thus spuriously contributing to the calculation of the total energy curvature~\cite{Cococcioni:2005}, it is important to maintain the Hubbard potential unchanged (and equal to its self-consistent 
unperturbed value) during the iterative solution of linear-response equations~\cite{Timrov:2018}, as was discussed above. This is the reason why no Hubbard response potential is included on the right-hand side of Sternheimer equations~\eqref{eq:KS_lin_eq_q}, \eqref{eq:LRKSeq_lp} and \eqref{eq:LRKSeq_lp_us}.

\begin{figure}[t]
\begin{center}
   \includegraphics[width=0.47\textwidth]{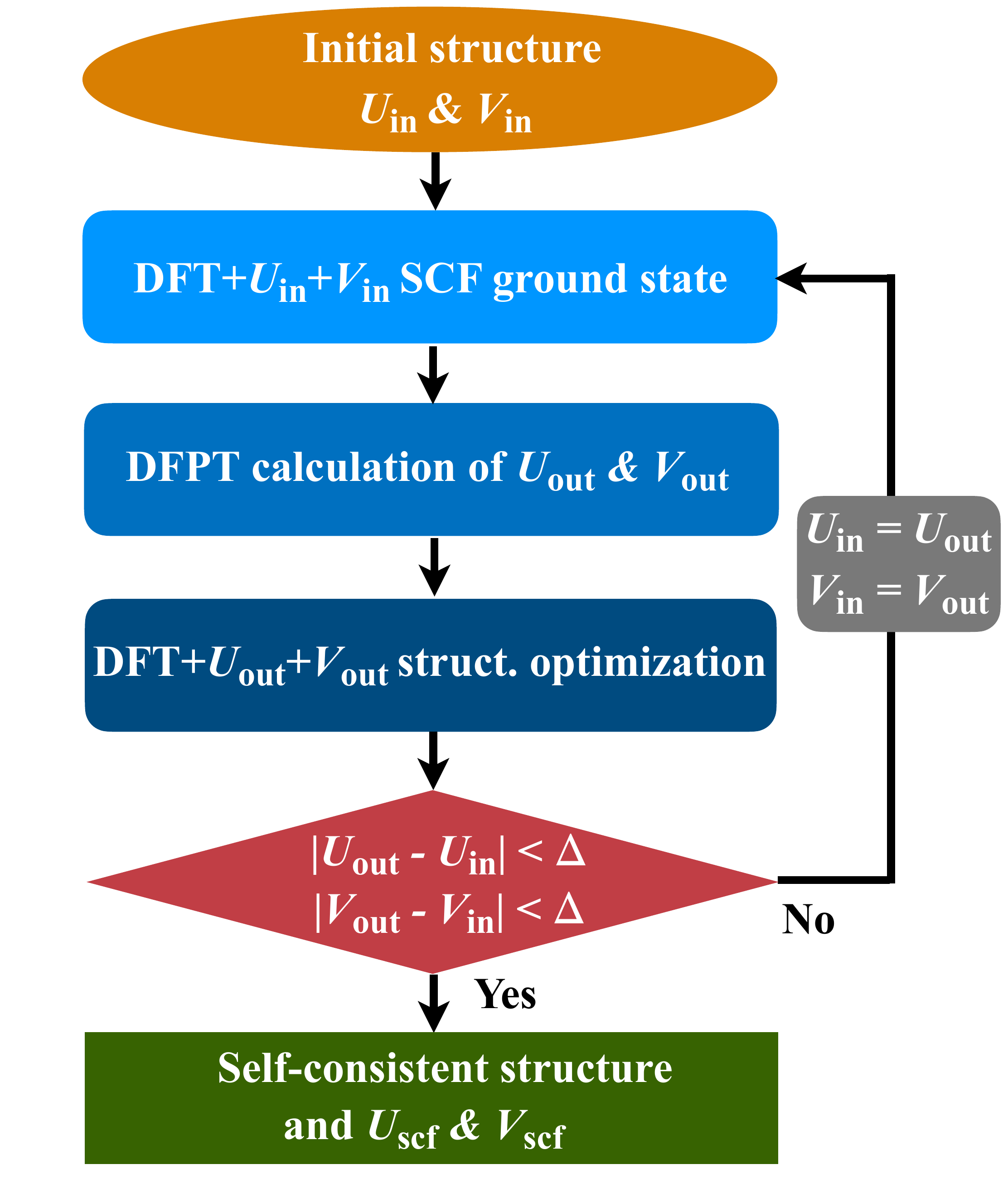}
   \caption{A protocol for the calculation of self-consistent Hubbard parameters $U_\mathrm{scf}$ (on-site) and $V_\mathrm{scf}$ (inter-site).}
\label{fig:UV_scf_scheme}
\end{center}
\end{figure}

The procedure described up to this point obtains the self-consistent values of $U$ and $V$ for a given crystal structure. When structural optimization is necessary it is also important to obtain consistency of the final Hubbard parameters with the crystal geometry as well, so one can envision an overall iterative procedure, as illustrated by the flowchart diagram in Fig.~\ref{fig:UV_scf_scheme}. The importance of the consistency of the Hubbard parameters with the crystal structure has already been highlighted in some recent work involving the authors~\cite{Cococcioni:2019, Ricca:2019, Ricca:2020}. In fact, the evaluation of the Hellmann-Feynman forces and stresses contains Pulay (Hubbard) terms that are computed at a fixed geometry and by assuming that $U$ and $V$ do not depend on the atomic positions and on the strain. However, as shown in Ref.~\cite{Kulik:2011b}, Hubbard parameters can depend on atomic positions, and this should be taken into account when computing $U$ and $V$ self-consistently with the crystal structure. So, the workflow of Fig.~\ref{fig:UV_scf_scheme} drives the system to a global minimum also with respect to the changes in the Hubbard parameters (this will be discussed in more detail elsewhere).

In summary, the procedure to compute $U_\mathrm{scf}$ and $V_\mathrm{scf}$ is as follows:

\begin{itemize}

\item {\it Step 1.} Set up the initial crystal structure for the system of interest and choose the initial values for $U$ and $V$ (possibly zero);

\item {\it Step 2.} Perform a self-consistent field (SCF) ground state calculation using DFT+$U$+$V$ with the current input values of Hubbard parameters (i.e., $U_\mathrm{in}$ and $V_\mathrm{in}$);

\item {\it Step 3.} Perform a DFPT calculation to determine the new values of the Hubbard parameters ($U_\mathrm{out}$ and $V_\mathrm{out}$); 

\item {\it Step 4.} Perform a structural optimization using DFT+$U$+$V$ with the current values of $U$ and $V$ (i.e., $U_\mathrm{out}$ and $V_\mathrm{out}$);

\item {\it Step 5.} Return to step 2 with updated values of the Hubbard parameters (i.e., $U_\mathrm{in} = U_\mathrm{out}$ and $V_\mathrm{in} = V_\mathrm{out}$) and iterate the procedure until the point when the variations of the Hubbard parameters [see Eqs.~\eqref{eq:U_conv} and \eqref{eq:V_conv}] and of the crystal structure are both within fixed thresholds. 

\end{itemize}

In view of the fact that the DFT+$U$(+$V$) electronic ground state is often qualitatively different from the DFT one, it is usually a good idea to start the self-consistent loop in Fig.~\ref{fig:UV_scf_scheme} from finite $U_\mathrm{in}$ and $V_\mathrm{in}$, to avoid oscillations in their value during the initial steps of their calculations. Since $U$ and $V$ depend on the Hubbard manifold on which they act, a reasonable guess of $U_\mathrm{in}$ and $V_\mathrm{in}$ can only come from a calculation done for a similar material and the same pseudopotential. Of course, it is always possible to start with $U_\mathrm{in}$ and $V_\mathrm{in}$ set to 0 (as done here for demonstration purposes). In systems with multiple (meta)stable states (e.g., spin-polarized molecules) the converged values of the Hubbard parameters will be different (albeit often only slightly) depending on the specific minimum that is reached. We also note in passing that the actual minimum that is reached will depend more on the symmetry of the system, the initial magnetization or the starting atomic occupations than the actual chosen values of $U_\mathrm{in}$ and $V_\mathrm{in}$~\cite{Meredig:2010}.

The current wokflow for the calculation of $U_\mathrm{scf}$ and $V_\mathrm{scf}$ can be easily implemented and automatized. Since the Hubbard parameters depend on many computational details (chemical composition, choice of localized functions for projectors, xc functional, pseudopotentials, oxidation state, etc.), having a fully automated workflow for calculation of $U_\mathrm{scf}$ and $V_\mathrm{scf}$ is indispensable for the high-throughput materials' screening.

\section{Technical details}
\label{sec:technical_details}

In order to exemplify and benchmark the application of the DFPT approach described here we chose MnPO$_4$ and LiMnPO$_4$ as test cases. In this section we review the technical settings of these calculations. The DFPT approach for calculations of Hubbard parameters with US PPs and PAW discussed here has been implemented in the \textsc{Quantum ESPRESSO} distribution~\cite{Giannozzi:2009, Giannozzi:2017, Giannozzi:2020} and has been released starting with version 6.6~\cite{QuantumESPRESSO:website}.

All calculations are performed using the plane-wave (PW) pseudopotential method and the xc functional is constructed using spin-polarized GGA with the PBEsol prescription~\cite{Perdew:2008}. Pseudopotentials are chosen based on the SSSP library~1.1 (efficiency)~\cite{Prandini:2018, MaterialsCloud}: we have used US PPs and PAW~\cite{Timrov:Note:2020:PP} from the Pslibrary~0.3.1 and 1.0.0~\cite{Kucukbenli:2014, Dalcorso:2014} and the GBRV library~1.2 and 1.5~\cite{Garrity:2014}. The crystal structure is optimized at three levels of theory (DFT, DFT+$U$, and DFT+$U$+$V$) using the Broyden-Fletcher-Goldfarb-Shanno (BFGS) algorithm~\cite{Fletcher:1987}, with a convergence threshold for the total energy of $5 \times 10^{-6}$~Ry and for forces of $5 \times 10^{-5}$~Ry/Bohr. We used an antiferromagnetic ordering for both materials (labeled ``AF$_1$'' in Ref.~\cite{Cococcioni:2019}).
To construct the projectors $\varphi^I$ of the Hubbard manifold [see Eq.~\eqref{eq:Pm1m2us}] we have used atomic orbitals which are orthogonalized using L\"owdin's method~\cite{Lowdin:1950, Mayer:2002}; structural optimizations using DFT+$U$ and DFT+$U$+$V$ are performed using orthogonalized atomic orbitals as described in detail in Ref.~\cite{Timrov:2020b}. KS pseudo-wavefunctions and potentials are expanded in PWs up to a kinetic-energy cutoff of 90 and 1080~Ry, respectively, for structural optimization, and of 65 and 780~Ry, respectively, for calculation of Hubbard parameters and voltages. 

\begin{table*}[t]
 \begin{center}
  \begin{tabular}{ccccccc}
    \hline\hline
    \multirow{2}{*}{Method} & \parbox{2.2cm}{Hubbard} & \multicolumn{2}{c}{\parbox{3.5cm}{MnPO$_4$}} & & \multicolumn{2}{c}{\parbox{3.5cm}{LiMnPO$_4$}} \\ \cline{3-4} \cline{6-7}
                &      parameters         &       LR-cDFT       &     DFPT               & &        LR-cDFT       &      DFPT                \\ \hline
    DFT+$U$     &    $U$(Mn)              &       5.619         &     5.619              & &         5.073        &     5.072                \\ \hline 
    \multirow{5}{*}{DFT+$U$+$V$} & $U$(Mn)&       6.215         &     6.214              & &         5.401        &     5.400                \\
                &    $V$(Mn-O$_1$)        &       1.193         &     1.193              & &         0.638        &     0.638                \\
                &    $V$(Mn-O$_2$)        &       1.214         &     1.214              & &         0.907        &     0.907                \\
                &    $V$(Mn-O$_3$)        &       1.090         &     1.090              & &         0.926        &     0.926                \\
                &    $V$(Mn-O$_4$)        &       0.666         &     0.666              & &         0.539        &     0.539                \\
     \hline\hline
  \end{tabular}  
 \end{center}
\caption{Comparison of the Hubbard parameters $U$ and $V$ (in eV) computed using LR-cDFT and DFPT, for MnPO$_4$ and LiMnPO$_4$. The parameters were computed in a ``one-shot'' fashion (i.e. no self-consistency described in Fig.~\ref{fig:UV_scf_scheme}), using orthogonalized atomic orbitals, and starting from the DFT ground state on top of the DFT optimized geometry.}
\label{tab:Benchmark}
\end{table*}

The olivines MnPO$_4$ and LiMnPO$_4$ have an orthorhombic crystal structure; the theoretical values of the lattice parameters are compared with the experimental values in Sec.~\ref{sec:Voltages}. The Mn atoms are coordinated by six O atoms forming an octahedron (indicated as MnO$_6$) of which it occupies the center. The P atoms are instead at the center of PO$_4$ tetrahedra that they form with neighboring oxygens. The three-dimensional structure of the crystal can be understood as being based on a network of corner-sharing MnO$_6$ octahedra further linked by ``interstitial'' PO$_4$ tetrahedra that act as structural reinforcer [avoiding excessive volume variations upon Li (de-)insertion] and chemical stabilizers (useful to avoid oxygen escapes). Lithium ions reside within octahedral channels along the intermediate-length side ($b$) of the cell. The reader is referred to Ref.~\cite{Cococcioni:2019} for more details.

The DFPT calculations of Hubbard parameters (for production purposes) are performed using the uniform $\Gamma$-centered $\mathbf{k}$ and $\mathbf{q}$ point meshes of size $3 \times 4 \times 5$ and $1 \times 2 \times 3$, respectively, which give an accuracy of 0.01~eV for the computed values of $U$ and $V$. Unit cells with four formula units are used, containing 24 atoms in the case of MnPO$_4$ and 28 atoms in the case of LiMnPO$_4$~\cite{Cococcioni:2019}. The linear-response KS equations~\eqref{eq:LRKSeq_lp_us} are solved using the conjugate-gradient algorithm~\cite{Payne:1992} and the mixing scheme of Ref.~\cite{Johnson:1988} for the response potential to speed up convergence.

The DFPT calculations of Hubbard parameters (for benchmarking purposes) are performed using uniform $\Gamma$-centered $\mathbf{k}$ and $\mathbf{q}$ point meshes of size $3 \times 4 \times 6$ and $1 \times 2 \times 3$, respectively, and the same unit cells as described above. The equivalent LR-cDFT calculations of $U$ and $V$ were performed using finite differences on $1 \times 2 \times 3$ supercells (resulting in 144 and 168 atoms for MnPO$_4$ and LiMnPO$_4$, respectively) whose Brillouin zone was sampled using a correspondingly coarser $\Gamma$-centered $3 \times 2 \times 2$ $\mathbf{k}$ point mesh. For LR-cDFT, the strength of the perturbation is chosen to be $\pm 0.05$~eV (see Ref.~\cite{Cococcioni:2005} for more details).

The data used to produce the results of this work are available in the Materials Cloud Archive~\cite{MaterialsCloudArchive2020}.

\section{Results and discussion}
\label{sec:results}

\subsection{Benchmark of DFPT versus LR-cDFT}
\label{sec:Benchmark}

In this section we present a validation of the analytical formulas for DFPT presented in Sec.~\ref{ucalc} and their numerical implementation, in the framework of US PPs and PAW. For this purpose, we make a comparison with the calculations performed using the well-established LR-cDFT approach of Ref.~\cite{Cococcioni:2005}.

As described in Sec.~\ref{sec:technical_details}, DFPT and LR-cDFT calculations are set up under equivalent conditions so that they ought to give the same values of $U$ and $V$. For more details about the equivalence between these two methods we refer the reader to Ref.~\cite{Timrov:2018}. Table~\ref{tab:Benchmark} shows the comparison of Hubbard parameters computed from these two approaches for MnPO$_4$ and LiMnPO$_4$. The calculations are performed in a ``one-shot'' fashion, i.e. without performing the self-consistency loop as described in Sec.~\ref{sec:protocol} -- our goal here is not to compute $U$ and $V$ self-consistently (this will be discussed in Sec.~\ref{sec:SCF_calc}) but to validate the extension of the DFPT formalism to US PPs and PAW. DFPT and LR-cDFT calculations are performed starting from the DFT ground state and using the geometry that was optimized also at the DFT level (this is a typical starting point in the protocol described in Sec.~\ref{sec:protocol}).

\begin{figure*}[t]
\begin{center}
   \subfloat[]{\includegraphics[width=0.42\textwidth]{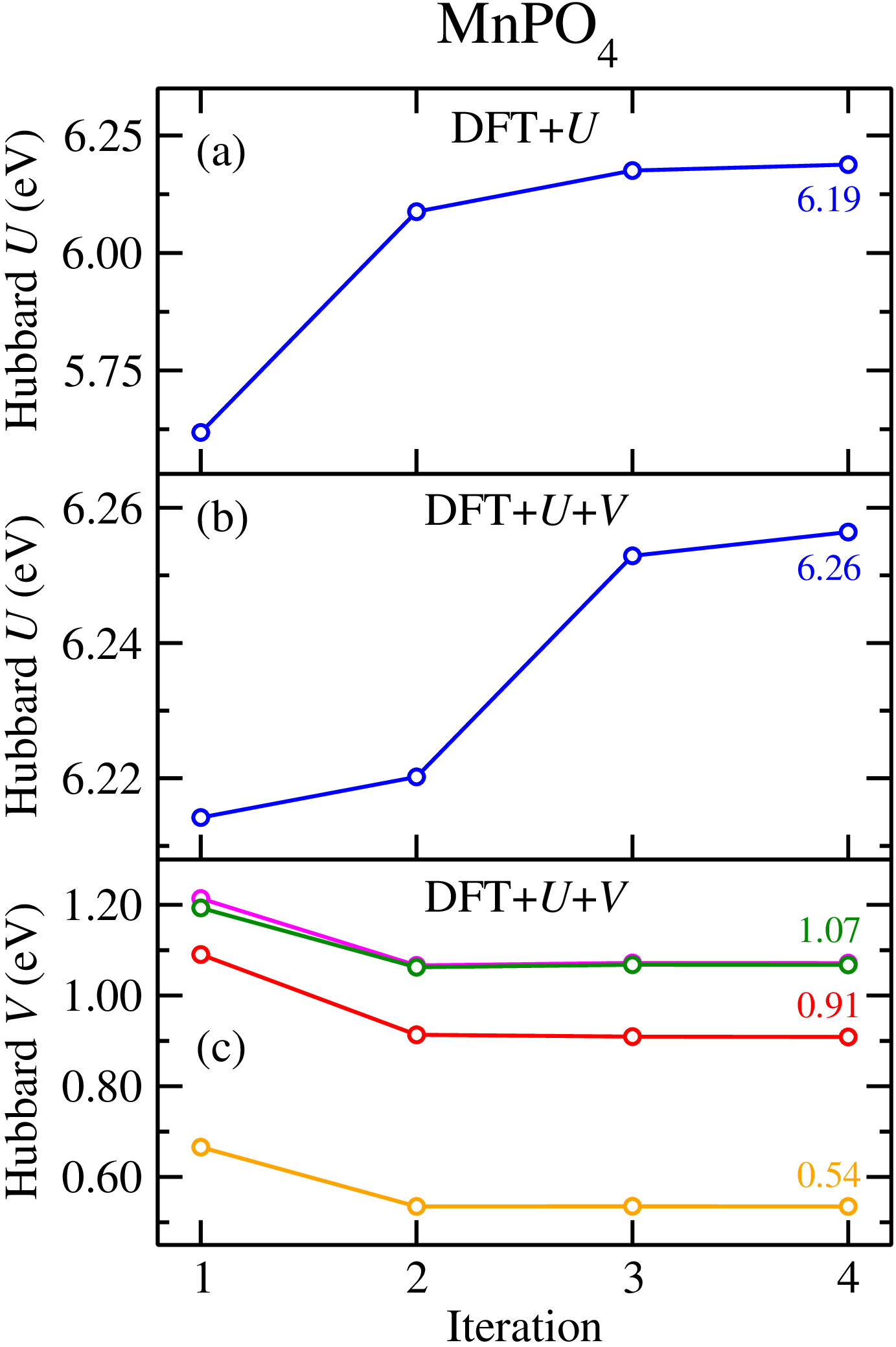}}
   \hspace{0.8cm}
   \subfloat[]{\includegraphics[width=0.42\linewidth]{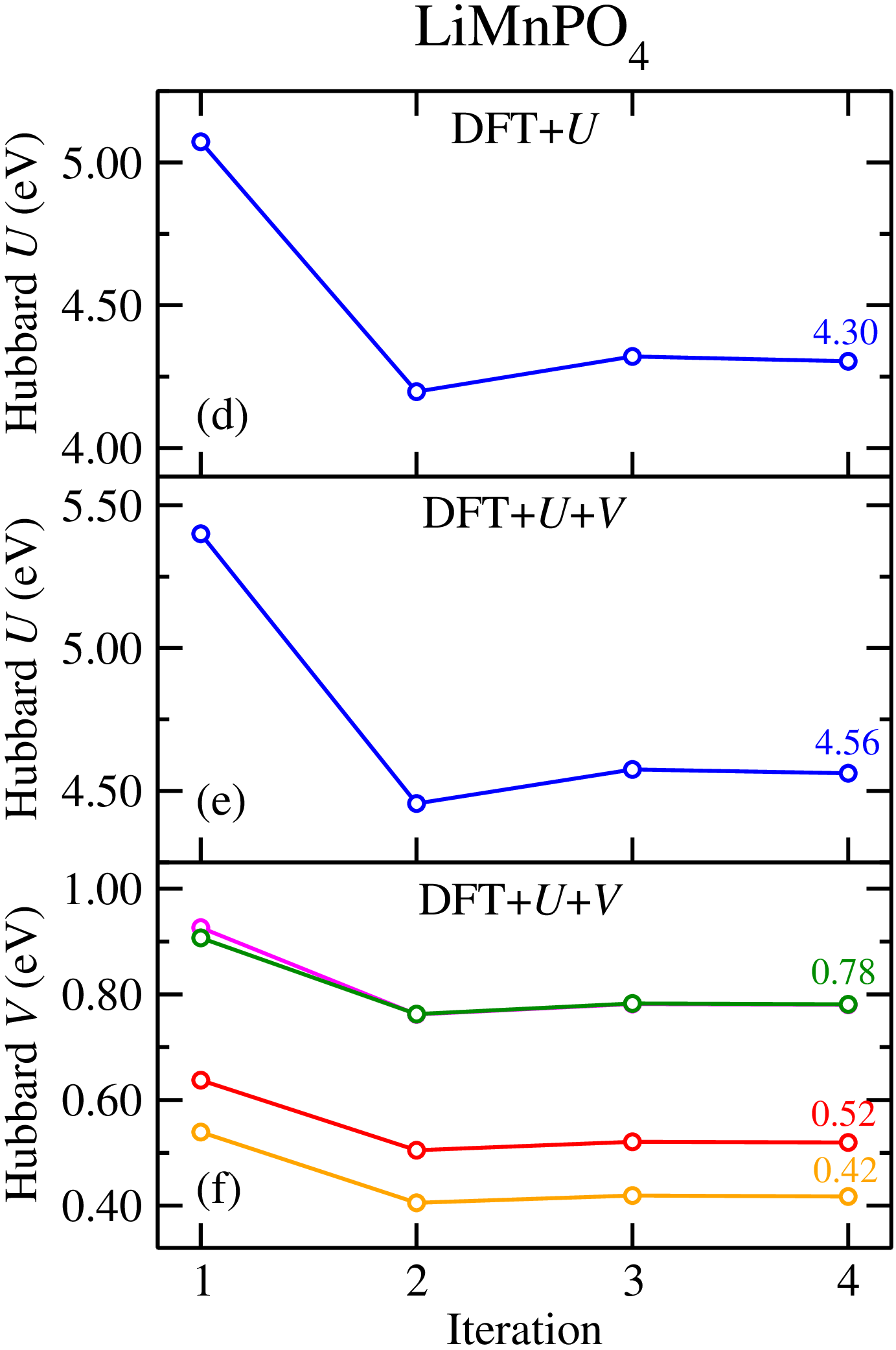}}
   \caption{Convergence of the Hubbard parameters $U$ and $V$ during their self-consistent evaluation from DFPT calculations based on DFT+$U$ and DFT+$U$+$V$. Results are presented for both MnPO$_4$ (left column), and LiMnPO$_4$ (right column). On-site $U$ refers to Mn($3d$) states while inter-site $V$'s are used between nearest-neighbor pairs of Mn($3d$) and O($2p$) states (each Mn being coordinated by four inequivalent O neighbors). All calculations are based on a basis set of orthogonalized atomic orbitals for the Hubbard manifold.}
\label{fig:HubbardParameters}
\end{center}
\end{figure*}

As can be seen from Table~\ref{tab:Benchmark}, the agreement between Hubbard parameters $U$ and $V$ computed using DFPT and LR-cDFT is excellent for both cases, and in line with the benchmark for the NC PPs implementation discussed in Ref.~\cite{Timrov:2018}. The on-site $U$ for Mn($3d$) and the inter-site $V$ between Mn($3d$) and O$(2p)$ states agree within $10^{-3}$~eV. The level of agreement could be improved even further by reducing the numerical noise through tighter convergence parameters; however, such accuracy on Hubbard parameters is not needed for any practical application where one wants parameters converged to 0.01--0.1~eV. It is worth noting that in MnPO$_4$ and LiMnPO$_4$ there are four inequivalent types of oxygen atoms (we label them as O$_1$, O$_2$, O$_3$, and O$_4$), and this is the reason why we obtain four values of inter-site Hubbard $V$ parameters between those and the neighbor transition-metal center (Mn atom). For different pairs of atoms, the Hubbard $V$ parameters are different due to variation in inter-atomic distances and local chemical environment (electronic screening), reminding us of the importance of locally sensitive Hubbard parameters.

Even though the systems used here for benchmark purposes are insulating (even at the DFT level), the extension of DFPT to metals is needed for systems that are metallic (and possibly become insulating after the application of the Hubbard corrections). Such a benchmark will not be discussed here, but it has also been investigated in detail, finding excellent agreement between DFPT and LR-cDFT.

All the results discussed above lead us to conclude that the DFPT formalism is correctly extended to US PPs and PAW, and it can be reliably used  for the calculations of both on-site $U$ and inter-site $V$. Now we can proceed to the self-consistent evaluation of Hubbard parameters using the protocol described in Sec.~\ref{sec:protocol}.

\subsection{Self-consistent calculation of the Hubbard parameters}
\label{sec:SCF_calc}

In this section we showcase the self-consistent evaluation of Hubbard parameters $U$ and $V$ for MnPO$_4$ and LiMnPO$_4$ using the workflow described in Sec.~\ref{sec:protocol}. The results are shown in Fig.~\ref{fig:HubbardParameters}.

We can see that after just a few iterations (typically 2 or 3) the Hubbard parameters are already within 0.1~eV from the converged value. We remind the reader that each iteration requires a structural optimization, a ground state calculation, and a calculation of Hubbard parameters using DFPT. The first iteration is based on an uncorrected DFT calculation ($U=V=0$) of the equilibrium structure, while in all subsequent steps the DFPT evaluation of the Hubbard parameters is based on DFT+$U$ or DFT+$U$+$V$ calculations with $U$ and $V$ determined in the previous step.

As can be seen in Fig.~\ref{fig:HubbardParameters}, the variations of both $U$ and $V$ are in general non-monotonic, and the converged values can be either higher or lower than their values obtained at the first iteration, i.e. obtained from the uncorrected ground state. More specifically, the converged $U$ for LiMnPO$_4$ is about 1~eV smaller than its value at the first iteration, while the variation for MnPO$_4$ is less pronounced (especially with DFT+$U$+$V$). On the other hand, Hubbard $V$ presents much smaller variations -- typically within 0.2~eV of their value at the first iteration -- and, in addition, $V$ for different pairs of Mn and O atoms typically show a similar convergence trend. 

The structural optimization included in the self-consistent procedure (see Fig.~\ref{fig:UV_scf_scheme}) is of crucial importance for determining the final values of the Hubbard parameters. This certainly enhances the variation of $U$ and $V$ during the self-consistent cycle due to their sensitivity to the local chemical environment of the Hubbard atoms~\cite{Kulik:2011b} and to the consequent dependence on the interatomic distances, lattice parameters, cell geometry, and symmetry. As can be seen in Figs.~\ref{fig:HubbardParameters}~(c) and \ref{fig:HubbardParameters}~(f), the Hubbard $V$ parameters are very similar for some couples of Mn and O atoms (see the magenta and green lines) -- this is due to the fact that interatomic distances (after the structural optimization) are very similar for these couples. We argue that including the structural optimization step is crucial to obtaining Hubbard $U$ and $V$ which are fully consistent with the crystal structure.

The self-consistent Hubbard parameters $U_\mathrm{scf}$ and $V_\mathrm{scf}$ are indicated in Figs.~\ref{fig:HubbardParameters}~(a)--\ref{fig:HubbardParameters}~(f) [at iteration 4], and these values will be used in the next section for the discussion of structural properties of MnPO$_4$ and LiMnPO$_4$, and for the evaluation of the voltage of a Li$_x$MnPO$_4$-based cathode versus a standard Li/Li$^+$ reference.

\subsection{Structural properties and energetics}
\label{sec:Voltages}

The detailed investigation of structural, electronic, and magnetic properties of MnPO$_4$ and LiMnPO$_4$ was already performed in great detail by some of us in a previous study~\cite{Cococcioni:2019}. Here, we will discuss some improvements that we obtained thanks mainly to a more accurate evaluation of the Hubbard parameters achieved with the DFPT implementation, and a more consistent way to perform the structural optimization.

In fact, in Ref.~\cite{Cococcioni:2019} the calculation of Hubbard parameters was performed using LR-cDFT with orthogonalized atomic orbitals, while structural optimizations were performed using non-orthogonalized atomic orbitals, due to the lack of implementation of Hubbard forces and stresses with orthogonalized atomic orbitals. Although good agreement with experiments was obtained for various properties using the DFT+$U$+$V$ functional, the use of different Hubbard manifolds for the calculation of $U$ and $V$ on the one hand and for the structural optimization on the other hand is not completely justified. Here, instead, we make use of a recent extension and implementation of Hubbard forces and stresses based on orthogonalized atomic orbitals~\cite{Timrov:2020b}. Moreover, instead of LR-cDFT as in Ref.~\cite{Cococcioni:2019}, we use DFPT for the calculation of the Hubbard $U$ and $V$ parameters using orthogonalized atomic orbitals that, while equivalent to LR-cDFT (see Sec.~\ref{sec:Benchmark} and Ref.~\cite{Timrov:2018}), offer a better control of the numerical accuracy and the convergence of the Hubbard parameters. While the above mentioned advancements in the computational procedure are certainly relevant, it is important to note that the differences between the present results and those of Ref. \cite{Cococcioni:2019} are also due to the use of different pseudopotentials and the consequent adoption of different computational parameters.

\begin{table}[t]
 \begin{center}
  \begin{tabular}{cccccc}
    \hline\hline
    Material    & LP          &   DFT  &  DFT+$U$ & DFT+$U$+$V$ &     Expt.     \\ \hline
    LiMnPO$_4$  & $a$         &  19.61 &   19.80  &    19.79    &  19.76$^\mathrm{a}$, 19.71$^\mathrm{b}$ \\
                & $b/a$       &   0.58 &    0.58  &     0.58    &     0.58$^\mathrm{a,b}$ \\
                & $c/a$       &   0.46 &    0.45  &     0.45    &     0.45$^\mathrm{a,b}$ \\ \hline
    MnPO$_4$    & $a$         &  18.40 &   18.52  &    18.46    &     18.31$^\mathrm{b}$ \\
                & $b/a$       &   0.61 &    0.61  &     0.61    &     0.61$^\mathrm{b}$  \\
                & $c/a$       &   0.50 &    0.50  &     0.50    &     0.49$^\mathrm{b}$  \\
     \hline\hline
  \end{tabular}
 \end{center}
\caption{The equilibrium lattice parameters (LP) $a$, $b$, and $c$ (in Bohr) of LiMnPO$_4$ and MnPO$_4$ computed using DFT, DFT+$U$, and DFT+$U$+$V$ with $U_\mathrm{scf}$ and $V_\mathrm{scf}$ determined as explained in Sec.~\ref{sec:SCF_calc}. The experimental data is from Ref.~\cite{Xiao:2010} (superscript $a$) and Ref.~\cite{Nie:2010} (superscript $b$).}
\label{tab:Lattice_parameters}
\end{table}

Table~\ref{tab:Lattice_parameters} shows the lattice parameters of LiMnPO$_4$ and MnPO$_4$ using DFT, DFT+$U$, and DFT+$U$+$V$, based on our refined procedure. For LiMnPO$_4$, DFT and DFT+$U$+$V$ results are in very good agreement with Ref.~\cite{Cococcioni:2019}, while our DFT+$U$ calculations give $a=19.80$~Bohr and in Ref.~\cite{Cococcioni:2019} DFT+$U$ gave $a=19.94$~Bohr. Thus, our DFT+$U$ $a$ is much closer to our DFT+$U$+$V$ $a$, which is likely due to the fact that in our calculations the structure is optimized more accurately including the effect of changes in the orbital overlap matrix and respective changes in Hubbard parameters. For MnPO$_4$, the difference between our lattice parameters and those of Ref.~\cite{Cococcioni:2019} is more pronounced. We now obtain a much better agreement with experimental lattice parameters (for both DFT+$U$ and DFT+$U$+$V$) than was achieved before, although, surprisingly, we still find standard DFT predictions to be the closest to the experimental data. 

\begin{figure}[t]
\begin{center}
   \includegraphics[width=0.37\textwidth]{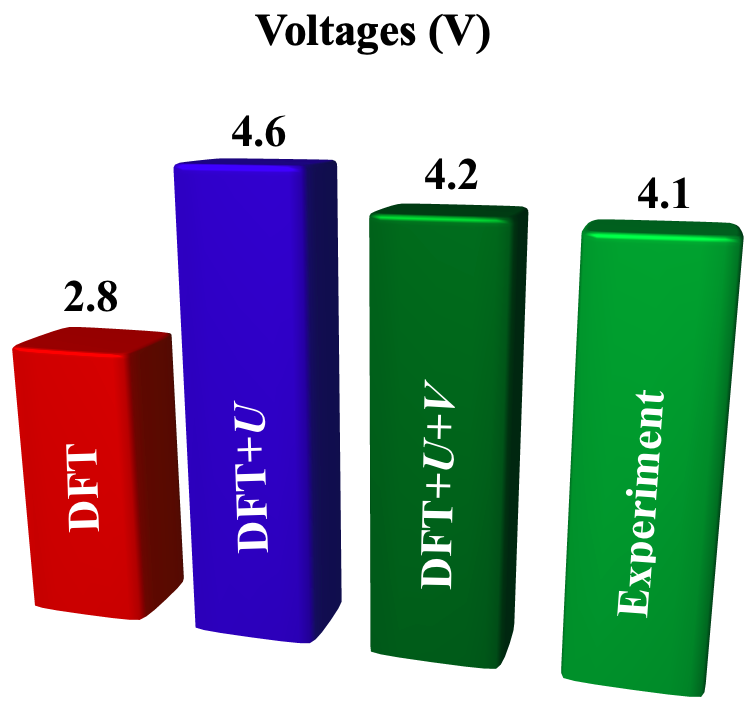}
\caption{Voltages $\phi$ (in V) in Li$_x$MnPO$_4$ (vs. Li/Li$^+$) computed using DFT, DFT+$U$, and DFT+$U$+$V$ with $U_\mathrm{scf}$ and $V_\mathrm{scf}$ determined in Sec.~\ref{sec:SCF_calc}. The experimental data is from Ref.~\cite{Padhi:1997}.}
\label{fig:Voltages}
\end{center}
\end{figure}

Figure~\ref{fig:Voltages} shows a comparison of voltages as measured in the experiments and computed in this work using the fundamental thermodynamic definition~\cite{Aydinol:1997}:
\begin{equation}
    \phi = - \frac{1}{e} \left[ E_{\mathrm{DFT}+U+V}^{\mathrm{LiMnPO}_4} - E_{\mathrm{DFT}+U+V}^{\mathrm{MnPO}_4} - E_{\mathrm{DFT}}^{\mathrm{Li}} \right] ,
    \label{eq:voltage}
\end{equation}
where $\phi$ is the voltage, $e$ is the electronic charge, $E_{\mathrm{DFT}+U+V}^{\mathrm{LiMnPO}_4}$ and $E_{\mathrm{DFT}+U+V}^{\mathrm{MnPO}_4}$ are the total energies of LiMnPO$_4$ and MnPO$_4$ computed at the DFT+$U$+$V$ level of theory using self-consistent values of $U$ and $V$ (see Sec.~\ref{sec:SCF_calc}), and $E_{\mathrm{DFT}}^{\mathrm{Li}}$ is the DFT total energy of bulk Li. It is important to note that the total energies entering in Eq.~\eqref{eq:voltage} are referred to the same amount of formula units and $E_{\mathrm{DFT}}^{\mathrm{Li}}$ represents the total energy of an equivalent number of Li atoms in bulk Li. We find that our DFT and DFT+$U$+$V$ results are consistent with those of Ref.~\cite{Cococcioni:2019}, while our DFT+$U$ voltage is 4.6~V versus 5.1~V of Ref.~\cite{Cococcioni:2019}. This latter result is a consequence of the difference in the values of $U$, and of the consistent calculation of forces and stresses using the orthogonalized atomic manifold that has significantly refined the prediction of the equilibrium crystal structure in this work.

\section{Conclusions}
\label{sec:Conclusions}

We have presented a generalization of density-functional perturbation theory to the linear-response calculation of both on-site $U$ and inter-site $V$ Hubbard parameters, using ultrasoft pseudopotentials and the projector-augmented wave method. This formalism is applicable to both insulating and metallic ground states, and it does not require the use of computationally expensive supercells with localized perturbations; instead, it is based on calculating the linear response of systems to monochromatic (i.e., wavelength-specific) perturbations inexpensively applied to primitive unit cells~\cite{Timrov:2018}. The work also discusses and applies an iterative procedure to compute the Hubbard parameters in full consistency with both the electronic and crystal structures. Moreover, the level of consistency is improved compared to previous works by the application of a recently developed algorithm~\cite{Timrov:2020b} for the calculation of Hubbard forces and stresses using orthogonalized atomic orbitals as a basis set for the Hubbard manifold.

The new extension of DFPT to US PPs and PAW has been benchmarked against the supercell LR-cDFT approach for LiMnPO$_4$ and MnPO$_4$: the fully lithiated and delithiated members of a cathode material for Li-ion batteries. An excellent agreement is achieved for the Hubbard parameters $U$ and $V$ between the two equivalent techniques, which validates the correctness of the derivation and implementation of the present extension of the DFPT approach to the calculation of Hubbard parameters. The equilibrium lattice parameters and average voltages computed with the DFT+$U$+$V$ approach are in good agreement with experiments, as in Ref.~\cite{Cococcioni:2019}, while those obtained with DFT+$U$ are sensibly improved with respect to the previous study. This latter outcome is the result of the consistency achieved between Hubbard parameters and the crystal structure due to the calculation of atomic forces and stresses on a Hubbard manifold of orthogonalized atomic orbitals~\cite{Timrov:2020b}.

What are the reasons for the accuracy of DFT+$U$+$V$ in describing complex materials? This approach does not actually cure correlation shortcomings -- strong correlations are really challenging. What it does, very well, is to cure major shortcomings {\it in the energetics} that many strongly-correlated systems have, originating in strong self-interaction errors driven by very localized $d$ and $f$ electrons. In this respect, DFT+$U$ and DFT+$U$+$V$ correct self-interactions in similar ways to hybrid functionals~\cite{anisimov:1997}, but with a strength that is, thanks to the linear-response approach adopted, sensitive to the electronic-structure environment of the system at hand, and thus ultimately both accurate and efficient. Incidentally, for these reasons these Hubbard functionals can also be applied to molecular
systems~\cite{Kulik:2006, Kulik:2011} where there is only a single transition-metal atom, and thus no
multi-site correlations.

Finally, it is important to mention that the new framework based on DFPT is implemented in the open-source \QE \, distribution~\cite{Giannozzi:2009, Giannozzi:2017, Giannozzi:2020} and is publicly available to the community at large. This approach, simple and accurate, allows one to perform simulations for complex transition-metal and rare-earth compounds, which are challenging for standard DFT. Notably, it can be easily incorporated in high-throughput workflows for simulations of thousands of materials, which could even be used to machine learn Hubbard parameters, with straightforward applications for data-driven materials modeling. Ultimately, we believe that this approach allows reliable, robust, and computationally affordable simulations to understand, predict, design, and discover novel materials.

\section*{ACKNOWLEDGMENTS}

This research was supported by the Swiss National Science Foundation (SNSF), through grant 200021-179138, and its National Centre of Competence in Research (NCCR) MARVEL. M.C. acknowledges partial support from the EU-H2020 research and innovation programme under Grant Agreement No.~654360 within the framework of the NFFA Europe Transnational Access Activity. Computer time was provided by CSCS (Piz Daint) through project No.~s836, and by the italian supercomputing center CINECA through the ISCRA project ``BatMat".

\appendix

\section{DFT+$U$+$V$ using the PAW method}
\label{app:PAW}

In this appendix we discuss in more detail how DFT+$U$+$V$ is used within the PAW method and its similarities to the case when US PPs are used.

In the PAW method, the KS all-electron wavefunctions $\Psi^\circ_{v\mathbf{k}\sigma}$ are reconstructed from the KS pseudo-wavefunctions $\psi^\circ_{v\mathbf{k}\sigma}$ as~\cite{Blochl:1994}:
\begin{equation}
    \Psi^\circ_{v\mathbf{k}\sigma}(\mathbf{r}) = \psi^\circ_{v\mathbf{k}\sigma}(\mathbf{r}) + 
    \sum_{J\mu} \left[ \phi^J_\mu(\mathbf{r}) - \tilde{\phi}^J_\mu(\mathbf{r}) \right] \braket{\beta^J_\mu}{\psi^\circ_{v\mathbf{k}\sigma}} \,,
    \label{eq:PAW_psi_AE}
\end{equation}
where $\phi^J_\mu$ and $\tilde{\phi}^J_\mu$ are the all-electron and pseudo partial waves (coinciding outside augmentation spheres), respectively, obtained from the PAW PPs. A similar expression holds for the all-electron atomic wavefunctions $\Phi^I_{m}$ that are obtained from the atomic pseudo-wavefunctions $\varphi^I_{m}$ as:
\begin{equation}
    \Phi^I_{m}(\mathbf{r}) = \varphi^I_{m}(\mathbf{r}) + 
    \sum_{J\mu} \left[ \phi^J_\mu(\mathbf{r}) - \tilde{\phi}^J_\mu(\mathbf{r}) \right] \braket{\beta^J_\mu}{\varphi^I_{m}} \,.
    \label{eq:PAW_phi_AE}
\end{equation}
From these expressions it is easy to compute (keeping in mind that $\langle \beta^I_\mu | \tilde{\phi}^J_\nu \rangle = \delta_{IJ} \delta_{\mu\nu}$ within the augmentation spheres~\cite{Blochl:1994}) the scalar product between the all-electron KS and atomic wavefunctions:
\begin{eqnarray}
    \bra{\Psi^\circ_{v\mathbf{k}\sigma}} \Phi^I_{m}\rangle & = & 
    \bra{\psi^\circ_{v\mathbf{k}\sigma}} \varphi^I_m \rangle +
    \sum_{J\mu\nu} \braket{\psi^\circ_{v\mathbf{k}\sigma}}{\beta^J_\mu} \nonumber \\
    & & \times \left[ \braket{\phi^J_\mu}{\phi^J_\nu} - \braket{\tilde{\phi}^J_\mu}{\tilde{\phi}^J_\nu} \right] 
    \braket{\beta^J_\nu}{\varphi^I_m} \nonumber \\ [6pt]
    & = & \bra{\psi^\circ_{v\mathbf{k}\sigma}} \hat S \vert \varphi^I_m  \rangle \,,
    \label{eq:PAW_A_matrix_element}
\end{eqnarray}
where $\hat S$ is the overlap operator defined in Eq.~\eqref{eq:S_overlap}, with the coefficients $q^{\gamma(I)}_{\mu\nu}$ given by $\braket{\phi^I_\mu}{\phi^I_\nu} - \braket{\tilde{\phi}^I_\mu}{\tilde{\phi}^I_\nu}$. 

The generalized occupation matrices thus result:
\begin{eqnarray}
n^{I J \sigma}_{m_1 m_2} &=& \sum_{\mathbf{k}}^{\nks} \sum_v 
   \tilde{\theta}_{F, v\mathbf{k}\sigma} \, 
   \braket{\Psi^\circ_{v\mathbf{k}\sigma}}{\Phi^J_{m_2}}
   \braket{\Phi^I_{m_1}}{\Psi^\circ_{v\mathbf{k}\sigma}} \nonumber \\
   & = & \sum_{\mathbf{k}}^{\nks} \sum_v 
   \tilde{\theta}_{F, v\mathbf{k}\sigma} \, 
   \bra{\psi^\circ_{v\mathbf{k}\sigma}} \hat{P}^{JI}_{m_2 m_1}
   \ket{\psi^\circ_{v\mathbf{k}\sigma}} \,, 
\label{eq:occ_matrix_0_app2}
\end{eqnarray}
where the projector $\hat{P}^{JI}_{m_2 m_1}$ is expressed in exactly the same way as for US PPs [see Eq.~\eqref{eq:Pm1m2us}], but with $\hat{S}$ defined in terms of augmentation and projector functions from the PAW scheme.

%
%

\section{Bloch sums}
\label{app:Bloch_sums}

Similarly to Appendix~1 of Ref.~\cite{Timrov:2018}, we define here Bloch sums and inverse Bloch sums of projector functions as, respectively, 

\begin{equation}
\tilde{\beta}^s_{\mu, \mathbf{k}}(\mathbf{r}) = 
\nksinvv \sum_{\mathbf{R}_l}^{\nks} e^{i\mathbf{k}\cdot\mathbf{R}_l} \, 
\beta^s_\mu(\mathbf{r} - \mathbf{R}_l) \,,
\label{eq:beta_Bloch_sum}
\end{equation}
and 
\begin{equation}
\beta^s_\mu(\mathbf{r}-\mathbf{R}_l) = 
\nksinvv \sum_{\mathbf{k}}^{\nks} e^{-i\mathbf{k}\cdot\mathbf{R}_l} \, 
\tilde{\beta}^s_{\mu, \mathbf{k}}(\mathbf{r}) \,.
\label{eq:beta_Bloch_sum_inv}
\end{equation}
The functions $\tilde{\beta}^s_{\mu, \mathbf{k}}(\mathbf{r})$ are Bloch-like functions 
and, hence, they can be written as:
\begin{equation}
\tilde{\beta}^s_{\mu, \mathbf{k}}(\mathbf{r}) = 
\nksinvv \, e^{i\mathbf{k}\cdot\mathbf{r}} \, 
\bar{\beta}^s_{\mu, \mathbf{k}}(\mathbf{r}) \,,
\label{eq:beta_Bloch_function}
\end{equation}
where
\begin{equation}
\bar{\beta}^s_{\mu, \mathbf{k}}(\mathbf{r}) \equiv
e^{-i\mathbf{k}\cdot\boldsymbol{\tau}_s} \, \bar{\beta}^{\gamma(s)}_{\mu, \mathbf{k}}(\mathbf{r} - \boldsymbol{\tau}_s) \,.
\label{eq:beta_Bloch_periodic_part}
\end{equation}
The augmentation functions, $Q^{l,s}_{\mu\nu}(\mathbf{r}) \equiv 
Q^s_{\mu\nu}(\mathbf{r} - \mathbf{R}_l)$, can also be represented using Bloch sums and
inverse Bloch sums, respectively, as:
\begin{equation}
\tilde{Q}^s_{\mu\nu,\mathbf{k}}(\mathbf{r}) = 
\nksinvv \sum_{\mathbf{R}_l}^{\nks} e^{i\mathbf{k}\cdot\mathbf{R}_l} \, 
Q^s_{\mu\nu}(\mathbf{r}-\mathbf{R}_l) \,,
\label{eq:Q_Bloch_sum}
\end{equation}
and
\begin{equation}
Q^s_{\mu\nu}(\mathbf{r}-\mathbf{R}_l) = 
\nksinvv \sum_{\mathbf{k}}^{\nks} e^{-i\mathbf{k}\cdot\mathbf{R}_l} \, 
\tilde{Q}^s_{\mu\nu,\mathbf{k}}(\mathbf{r}) \,.
\label{eq:Q_Bloch_sum_inv}
\end{equation}
Similarly to $\tilde{\beta}^s_{\mu, \mathbf{k}}(\mathbf{r})$ [see Eq.~\eqref{eq:beta_Bloch_function}], 
the functions $\tilde{Q}^s_{\mu\nu,\mathbf{k}}(\mathbf{r})$ are Bloch-like functions, and, therefore,
they can be written as:
\begin{equation}
\tilde{Q}^s_{\mu\nu,\mathbf{k}}(\mathbf{r}) = 
\nksinvv \, e^{i\mathbf{k}\cdot\mathbf{r}} \, \bar{Q}^s_{\mu\nu,\mathbf{k}}(\mathbf{r}) \,,
\label{eq:Q_Bloch_function}
\end{equation}
where 
\begin{equation}
\bar{Q}^s_{\mu\nu,\mathbf{k}}(\mathbf{r}) \equiv 
e^{-i\mathbf{k}\cdot\boldsymbol{\tau}_s} \, 
\bar{Q}^{\gamma(s)}_{\mu\nu,\mathbf{k}}(\mathbf{r} - \boldsymbol{\tau}_s) \,.
\label{eq:Q_Bloch_periodic_part}
\end{equation}
%


\begin{thebibliography}{103}
\bibitem [{Tim({\natexlab{c}})}]{Timrov:Note:2020:USPP_aug}%
  \BibitemOpen
  \href@noop {} {} \bibinfo {note} {{The charge density,
  being physical, cannot be made smoother, while the pseudo-wavefunction can,
  provided their square is augmented back.}}\BibitemShut {Stop}%
\bibitem [{Tim({\natexlab{d}})}]{Timrov:Note:2018:circ}%
  \BibitemOpen
  \href@noop {} {} \bibinfo {note} {{B}y superscript
  ``$^\circ$'' we denote quantities which refer to the unperturbed ground state
  of the system.}\BibitemShut {Stop}%
\bibitem [{Tim({\natexlab{e}})}]{Timrov:Note:2020:nij}%
  \BibitemOpen
  \href@noop {} {} \bibinfo {note} {{It is important to note
  that the response of the non-diagonal (with respect to the atomic sites)
  components of the occupation matrix in Eq.~\eqref{eq:occ_matrix_0} is not
  needed, because these terms are not present in the definition of the response
  matrices $(\chi_0)_{IJ}$ and $(\chi)_{IJ}$ (see
  Eq.~\eqref{chi}).}}\BibitemShut {Stop}%
\bibitem [{Tim({\natexlab{g}})}]{Timrov:Note:2017:theta}%
  \BibitemOpen
  \href@noop {} {} \bibinfo {note} {{T}he $\theta$ function
  can be chosen to be any smooth function which approximates a step-like
  function, because it comes from the condition $\theta(\varepsilon) +
  \theta(-\varepsilon) = 1$ which is used in the derivation of the equations
  \cite{deGironcoli:1995}. For example, in Refs.~\cite{deGironcoli:1995,
  Baroni:2001} it is considered to be equal to $\tilde{\theta}$ which depends
  on the smearing technique which is used, or it can be considered to be a step
  function as in Ref.~\cite{Dalcorso:2001}. However, in this work $\theta$ is
  taken to be a reascaled complementary error function, as implemented in the
  latest versions of the \QE package, which makes the numerical stability of
  the calculations more robust.}\BibitemShut {Stop}%
\bibitem [{Tim({\natexlab{j}})}]{Timrov:Note:2020:PP}%
  \BibitemOpen
  \href@noop {} {} \bibinfo {note} {{We used the following
  pseudopotentials: \texttt{O.pbesol-n-rrkjpaw\_psl.0.1.UPF} (PAW),
  \texttt{mn\_pbesol\_v1.5.uspp.F.UPF} (US),
  \texttt{P.pbesol-n-rrkjpaw\_psl.1.0.0.UPF} (US), and
  \texttt{li\_pbesol\_v1.4.uspp.F.UPF} (US).}}\BibitemShut {Stop}%
\end{thebibliography}

%

\end{document}